\pdfoutput=1

\documentclass{sig-alternate-10pt}

\usepackage{array}
\usepackage{url}
\usepackage{amsmath}
\usepackage{subfigure}

\usepackage{verbatim}		

\usepackage{color,soul}
\usepackage{colortbl}

\begin{document}

\title{Guadalupe: a browser design for heterogeneous hardware}

\author{
Zhen Wang$^\dag$, Felix Xiaozhu Lin$^\dag$, Lin Zhong$^\dag$, and Mansoor Chishtie$^\ddag$
\vspace{1.5mm}\\
\begin{tabular}{c c}
\affaddr{$^\dag$Rice University, Houston, TX} & \affaddr{$^\ddag$Texas Instruments, Dallas, TX}\\
\end{tabular}
}
\maketitle

\graphicspath{{figs/}}

\begin{abstract}

Mobile systems are embracing heterogeneous architectures by getting more types of cores and more specialized cores, which allows applications to be faster and more efficient.
We aim at exploiting the hardware heterogeneity from the browser without requiring any changes to either the OS or the web applications. 
Our design, Guadalupe, can use hardware processing units with different degrees of capability for matched browser services.
It starts with a weak hardware unit, determines if and when a strong unit is needed, and seamlessly migrates to the strong one when necessary.
Guadalupe not only makes more computing resources available to mobile web browsing but also improves its energy proportionality.
Based on Chrome for Android and TI OMAP4, We provide a prototype browser implementation for resource loading and rendering.
Compared to Chrome for Android, we show that Guadalupe browser for rendering can increase other 3D application's frame rate by up to 767\% and save 4.7\% of the entire system's energy consumption.
More importantly, by using the two cases, we demonstrate that Guadalupe creates the great opportunity for many browser services to get better resource utilization and energy proportionality by exploiting hardware  heterogeneity.

\end{abstract}



\section{Introduction}\label{sec:intro}
By making the operating system and hardware \textit{transparent} to web application\footnote{In this work, we use \textit{web application} to refer to both more traditional, static web pages and more modern, interactive and dynamic ones.} developers,  a web browser has evolved into a powerful platform for content and application distribution. 
Recent development in hardware, especially mobile system hardware, however, challenges this key advantage of web applications as versus native applications. 

As integrated circuits are hitting the power wall, modern computer systems, from servers to smartphones, are embracing \textit{heterogeneity} in their hardware by adding processing units of various degrees of specialization and processing capability. 
First of all, the added processing units increase the computational resources available, allowing better performance for multiprocessing systems. 
Furthermore, with heterogeneity, a computer system can not only execute a task on hardware customized for it with much higher energy \textit{efficiency} but also match the hardware capability with the task workload for improved energy \textit{proportionality}.  
To exploit hardware heterogeneity, native application developers either directly use the APIs or library associated with a specialized hardware unit, e.g.,~\cite{renderscript}, or provide ``hints'' to the underpinning operating system (OS), e.g.,~\cite{lin2012asplos}.   

Requiring web applications to do the same will unfortunately break the much valued system and hardware transparency of the web. 
Therefore, in this work, we ask: \textit{can web applications leverage heterogeneous hardware transparently?} 
Our answer is a browser design called \textit{Guadalupe}.
Guadalupe recognizes two orthogonal dimensions of hardware heterogeneity: specialization and capability. 
It allows browser designers to define a \textit{mapping pod}, which is a set of browser functions that can be mapped onto a group of hardware units of similar specialization, or a \textit{hardware specialization group}.
After the static mapping, Guadalupe leverages a browser's run-time knowledge about web applications to identify the hardware unit with the suitable capability in the hardware specialization group for the mapping pod.
It starts the mapping pod on the weak unit of the group but timely migrates it to a stronger one by demand at the run time. 
Guadalupe provides an efficient optimization to reduce the performance and energy overhead of such migrations.

Guadalupe is a design point, or a small design region from a rather large design space for exploiting heterogeneous hardware in the browser. 
In this paper, we provide the design principles that help us derive Guadalupe and describe a prototype implementation of it based on the open-source Chromium browser. The prototype maps two key mapping pods, i.e., resource loading and rendering, to the two extremes of specialization, i.e., general-purpose processors and graphics accelerators, respectively. 
Using a tablet development system for OMAP4 mobile application processor (SoC or System-on-Chip) from Texas Instruments, we demonstrate how Guadalupe realizes the key performance and efficiency benefits of hardware heterogeneity. 
We show that Guadalupe browser for rendering can reduce the 3D accelerator usage by up to 75\% and frees it for potential 3D tasks from other applications. On emerging mobile systems where multiple applications can run concurrently, e.g., Microsoft Surface
and Samsung Galaxy Note,
the resources freed by Guadalupe can increase the other 3D application's frame rate by 18.5\% to 767\%.
At the same time, Guadalupe browser reduces the energy consumption of the entire system by 4.7\%.

With Guadalupe and its implementation, we make the following contributions:
\begin{itemize}
\item A set of design principles for exploiting hardware heterogeneity for web applications in a transparent manner.
\item A specific browser design, Guadalupe, that follows the principles in exploiting heterogeneous hardware.
\item An implementation of Guadalupe based on Chromium and TI OMAP4 mobile SoC. 
We experimentally show Guadalupe improves resource utilization and energy proportionality for web applications.
\end{itemize}

To the best of our knowledge, Guadalupe is the first to explore the mapping between available hardware resources, in particular heterogeneous ones, and the mapping pods.
Our effort is orthogonal to related work that incorporates more OS functions, e.g., \cite{moshchuk2010serviceos}, and embrace parallelism, e.g.,~\cite{badea2010hotpar, mai2012hotpar, meyerovich2010www}. 
As these proposals bring more tasks into the browser and extract parallel tasks from browser services, they provide new mapping pods to consider for heterogeneous hardware units and increase the potential benefits of Guadalupe.

In this work, we present Guadalupe in the context of mobile systems because mobile systems are the leading platform in embracing heterogeneous hardware. 
We do expect its design principles will be applicable to browsers on more powerful systems when the latter slowly though inevitably embrace specialized hardware units of various strength. 

The rest of the paper is organized as follows. Section~\ref{sec:background} introduces the background of heterogeneous architecture and browser internals.
Section~\ref{sec:case} exemplifies the benefit of exploiting hardware heterogeneity with two case studies.
Section~\ref{sec:design} describes the principles and the design of Guadalupe.
Section~\ref{sec:implementation} describes the prototype implementation of Guadalupe design.
Section~\ref{sec:evaluation} presents the evaluation of Guadalupe browser for the two cases.
Section~\ref{sec:related} discuss the related work.
Section~\ref{sec:conclusion} concludes the paper.

\section{Background}\label{sec:background}
The key objective of Guadalupe is to execute a mapping pod on the most suitable hardware unit. Therefore, we first provide the background for heterogeneous hardware architectures and browser internals.

\subsection{Heterogeneous Architecture}\label{sec:background_hetero}
As modern computing systems are often power and energy constrained, heterogeneous architecture has become a popular strategy for higher power efficiency.  
Mobile systems have been a leading platform in embracing this strategy. 

\begin{figure}[t!]
\centering 
\includegraphics[width=0.45\textwidth]{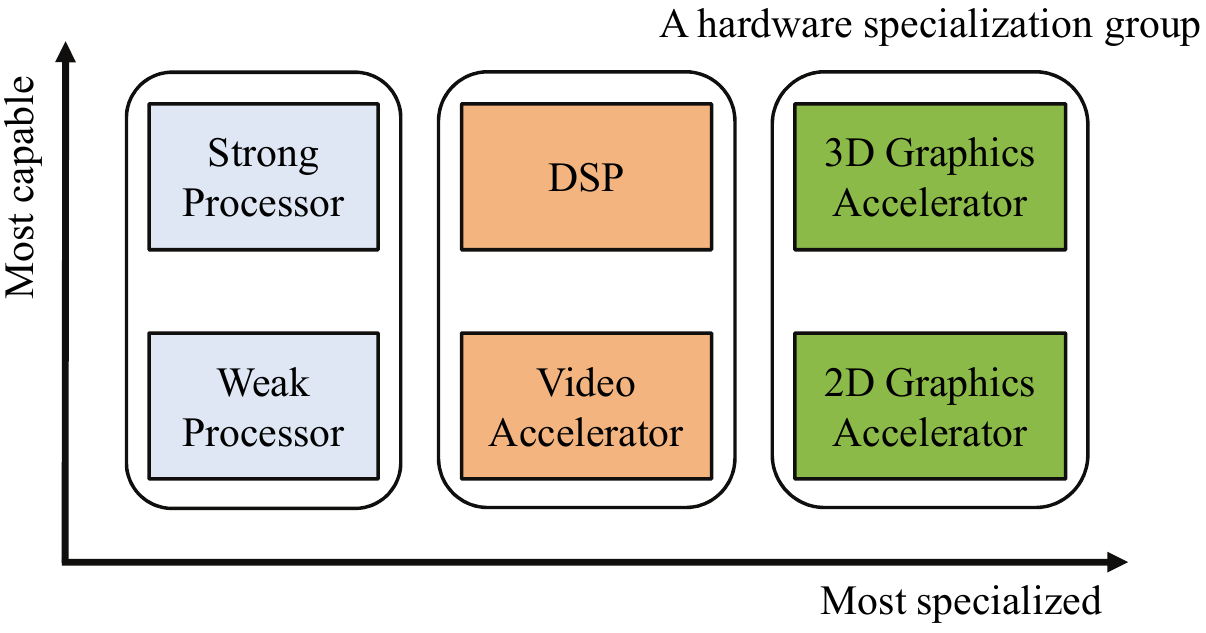}
\caption{Various heterogeneous hardware processing units with different specializations and capabilities on mobile SoCs}
\label{fig:hetero}
\end{figure}

Heterogeneity includes two orthogonal dimensions as illustrated by Figure~\ref{fig:hetero}. 
First, a heterogeneous architecture often employs processing units of various degrees of \textit{specialization}, from general-purpose processors like ARM cores to processors with special instruction set (ISA), e.g., digital signal processor (DSP),  to application-specific accelerators treated as I/O devices by the OS, e.g., graphics accelerator, as shown along the X axis of the figure. 
A specialized unit is often optimized for a specialized workload and can deliver the same performance with higher efficiency than general-purpose processors by orders of magnitude~\cite{hameed2010understanding}.  
For example, TI OMAP4470 mobile application processor~\cite{omap4470} has both ARM Cortex-A9 and M-3 cores, audio back end, DSP subsystem, image and video accelerator high-definition subsystem, display subsystem, face detect module, image subsystem, and graphics accelerator. 

Second, a heterogeneous architecture can employ processing units with different capabilities for the same specialization, as shown along the Y axis of Figure~\ref{fig:hetero}. 
More capable units have more functionalities and better performance, but they may also incur higher power consumption. 
We will use \textit{strong} and \textit{weak} to refer to more capable and less capable processing units of the same specialization in the rest of this paper, respectively. 
The dimension of capability is necessary because tasks of the same type and specialization may have a wide range of workload; and low-power and low-performance unit is necessary for light workload. 
For example, light-weight workload cannot fully utilize a powerful processor's architecture features, e.g., a deep pipeline, superscalar, speculative execution and large cache. 
Low-power processors can execute the light-weight workload with much higher efficiency~\cite{lin2012hotpower}. 
The emerging ARM big.LITTLE architecture~\cite{greenhalgh2011arm} attests to this strategy with general-purpose cores with different capabilities. 
TI OMAP4 also provides both Cortex-A9 and Cortex-M3 ARM cores as well as graphics accelerators of 2D and 3D capabilities.

The two dimensions of heterogeneity are dealt with differently, as will be discussed in Section~\ref{sec:design}: the browser designer statically maps the mapping pod to a hardware specialization; and Guadalupe dynamically determines the capability requirement. 
We will demonstrate the benefits of Guadalupe with the two extremes of the specialization dimension: general-purpose processors and graphics accelerators.

\subsection{Browser Internals}\label{sec:background_browser}
A browser is both an application and an ``OS''. 
A browser is an application running on the OS and needs many system resources such as CPU, graphics accelerators, memory, storage, network and I/O devices like touch screen. As an application, it accesses these resources through system calls provided by the underlying OS. 
However, a browser is also more than an application and starts to serve as a platform or ``OS'' for web applications, e.g., providing the interface to access hardware units~\cite{w3cdap} and enforcing the boundary between web applications~\cite{barth2008security}.

However, a browser has more knowledge about its web applications than a traditional OS has about its native applications. 
An OS knows very limited information of a native application and the information is passed from the native application through a well-defined interface consisted of system calls.
In contrast, the boundary between web applications and the browser is blurred.
The browser fetches source code of a web application, parses it and creates web application state inside the browser.
Therefore, the browser has almost full knowledge about web applications running on it, including their data structures and run-time behavior.
This gives the browser unique opportunities to determine the best hardware to execute tasks on behalf of web applications. In contrast, the developers of a native application often have to explicitly give ``hints'' to the OS to determine the best hardware for execution, e.g.,~\cite{anand2004mobisys, renderscript, lin2012asplos, patterson1995sosp}.

A browser represents the \textit{state} of a web application with several tree structures, namely, DOM tree, Render tree, and RenderLayer tree. 
The \textit{DOM tree} stores the web application content, e.g., text and images. 
The \textit{Render tree} has a one-to-one mapping to DOM tree's visible nodes and knows how to render them. 
The nodes in the Render tree are divided into several groups and each group corresponds to one render layer. 
The \textit{RenderLayer tree} ensures the correct rendering order among the render layers.

\begin{figure}[t!]
\centering 
\includegraphics[width=0.45\textwidth]{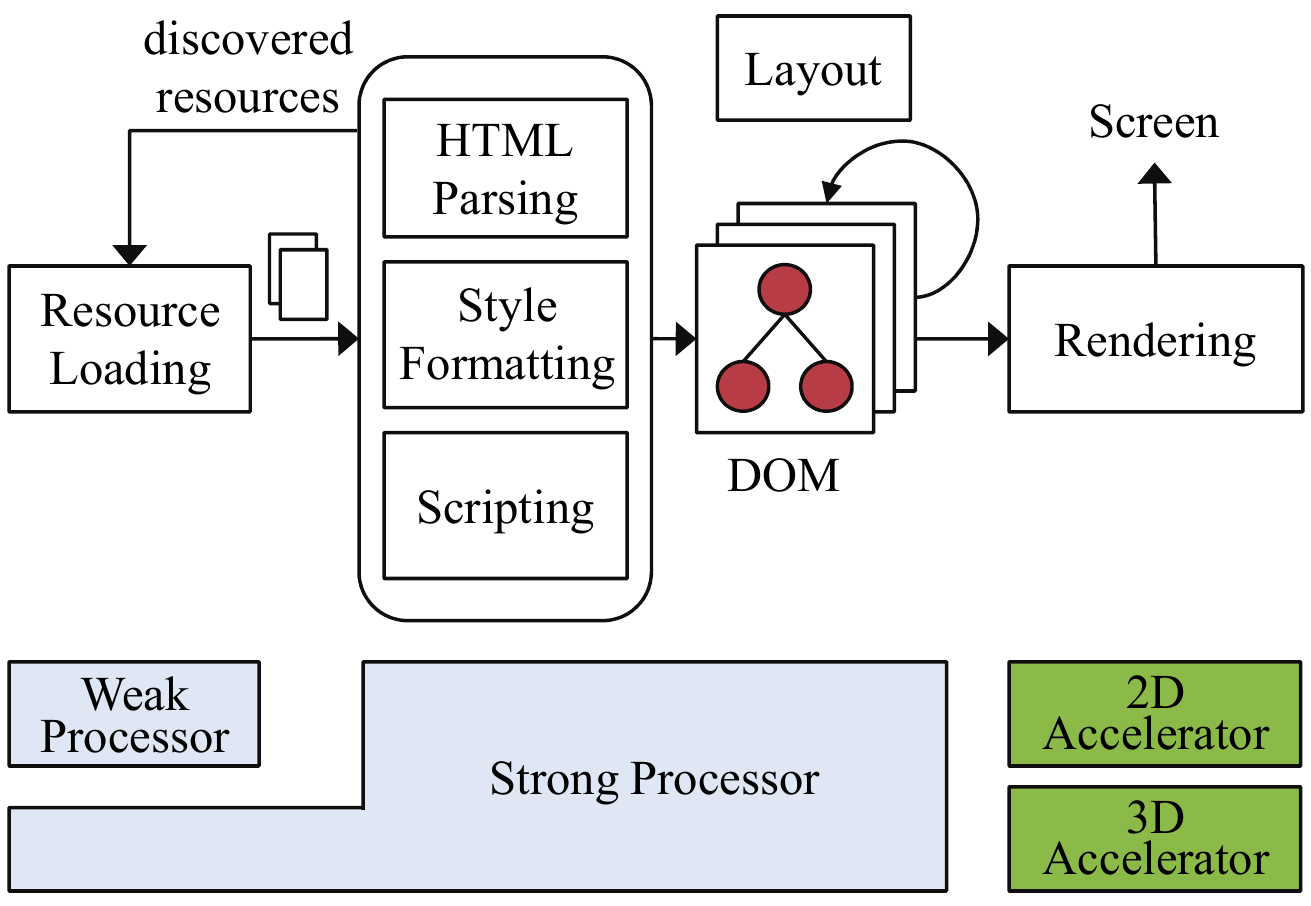}
\caption{Basic browser services on a heterogeneous SoC}
\label{fig:browser}
\end{figure}

A browser provides many services to a web application.
Each service is a collection of browser functions with similar semantics, i.e., working on the same type of objects and producing similar outputs.
A browser's functions have been naturally organized into six basic services, as shown in Figure~\ref{fig:browser}: resource loading, HTML parsing, style formatting, layout, scripting and rendering.
Resource loading fetches resource files referenced by a web page, e.g., HTML, CSS, JavaScript, and image files. 
HTML Parsing processes the HTML file and generates the DOM tree to represent the web page's content.
New resources may be discovered by the parsing service and the browser will load them accordingly.
Style formatting and layout calculate the styles and positions of the web page's content, respectively.
And they generate the Render tree and RenderLayer tree inside the browser.
Scripting executes JavaScript code to provide enhanced user interaction with the web by manipulating the web application's state, i.e., the tree structures in the browser.
Finally, rendering shows the web page's content onto the screen.

Apart from the six basic services, a browser also provides some add-on services, e.g., video decoding and image processing. 
With the evolution of the web and HTML standards, more and more add-on services will be added into the browser's functionalities.
We will discuss how a mapping pod is determined from those services and mapped onto a hardware specialization in Section~\ref{sec:design}.

\begin{figure*}[t!]
\centering
\begin{minipage}{.45\textwidth}
        \centering
        \includegraphics[width=1\textwidth]{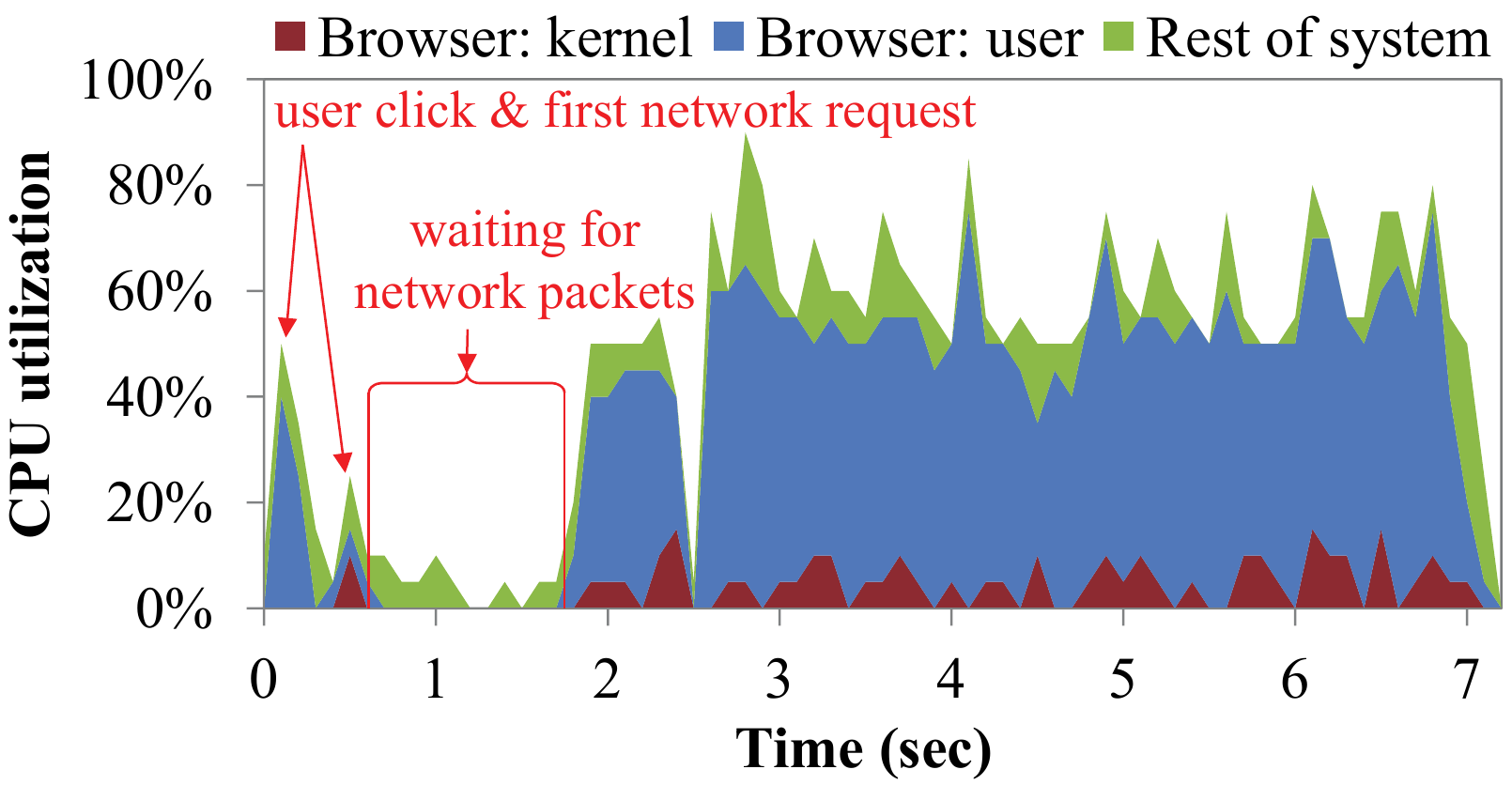}
        \caption{Measured CPU utilization when a browser is opening a CNN news page with 3G network on a smartphone. ``Browser: kernel'' and ``Browser: user'' correspond to the CPU utilization of the browser process in kernel space and user space, respectively.}
        \label{fig:loading}
\end{minipage}
\quad\quad\quad
\begin{minipage}{.45\textwidth}
        \centering
        \includegraphics[width=1\textwidth]{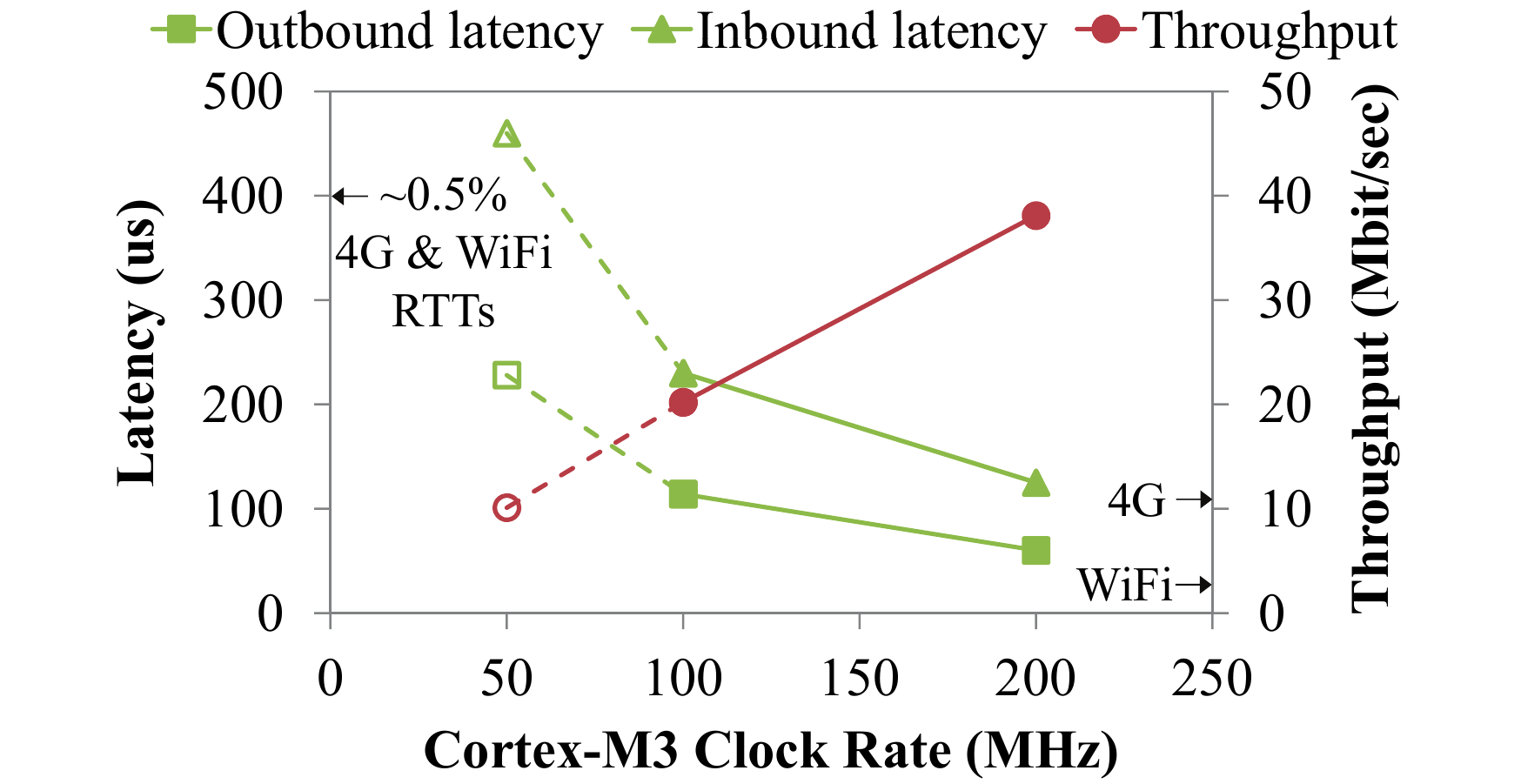}
        \caption{TCP loopback performance on Cortex-M3 of OMAP4.
        The solid markers are measured while the hollow ones are extrapolated.
        For comparison, typical wireless latency and bandwidth 
        \protect\cite{huang2012mobisys} are marked on y-axis.}
        \label{fig:resoure-loading-m3-performance}
\end{minipage}
\end{figure*}

\section{Case Studies}\label{sec:case}
Various services performed by the browser can greatly benefit from heterogeneous hardware, in terms of resource utilization and energy efficiency. We next use two cases, resource loading and rendering, to exemplify the benefits and lay out the facts that motivate the Guadalupe design.

\subsection{Resource Loading}\label{sec:loading}Resource loading feteches a resource given its URL. 
When the browser opens a web page, it first requests the main resource, which is usually an HTML document. 
After downloading and parsing the main resource, the browser can  usually discover more resources that are needed, e.g., CSS, JavaScript, and image files. 
They are called subresources. 
The browser will then fetech those subresources for additional content or page format and manipulation.

Resource loading can be parallel with other browser services beacuse of the browser's incremental rendering feature. 
Incremental rendering enables the browser to show the partially downloaded web page to the user while the browser is still loading more resources. 
For example, while the browser is loading subresources, the browser can layout the web page and render the partially downloaded web page onto the screen.

Resource loading is usually bounded by network latency.
Mobile browsers can take 2 seconds to get the first data packet of the main resource under 3G network~\cite{wang2012www}. 
During this period, the browser experiences light workloads, spending most of time blocking on the network IO.
Afterwards, while loading subresources, the browser starts to experience heavy workloads because subresource loading is parallel to other intensive browser services such as rendering.
Figure~\ref{fig:loading} shows the CPU utilization of a browser process in opening a web page.
During the early stage of page opening, i.e., before 1.8 sec, the browser incurs relatively low CPU usage, i.e., 10X lower than that of the later stage, waiting for the first a few packets.
Only after that, the browser starts to consume more CPU time, in parsing resource files and rendering web contents.


\textbf{TCP loopback micro-benchmark}.
In today's heterogeneous  SoCs,  such early-stage resource loading can be executed with typical weak processors with imperceivable performance loss.
We see evidence of this by measuring the performance of TCP/IP, 
the heart of resource loading, on a Cortex-M3 processor of OMAP4 SoC.

In the experiment, we employ the TCP loopback benchmark: the M3 core streams 1000-byte TCP packets to and from a \texttt{loopback} interface. With stressing processors and memory, TCP loopback is a widely accepted benchmark for network stack performance.
To develop the benchmark, we port \texttt{lwIP} \cite{dunkels2001lwip}, a lightweight yet full-fledged TCP/IP stack to Cortex-M3, 
bootstrapped by a preliminary version of our Kage kernel \cite{lin2012hotpower}.
We disable zero-copy to include real data movement overhead.
Other than that, we have not fine tuned the port due to time constraints.
As the Cortex-M3 on OMAP4 only has two possible clock rates, 
we extrapolate the measured results in order to show the performance trend.
Note that this limitation is specific to the OMAP4 platform and is not fundamental.


As shown in Figure \ref{fig:resoure-loading-m3-performance},  the TCP loopback benchmark implies that resource loading is able to achieve good performance on Cortex-M3.
Even with M3 running at 50 MHz, one fourth of its maximum clock rate, 
the network stack reaches a throughput of 10 Mbps, close to the typical 13 Mbps bandwidth of today's 4G network; 
with M3 running at 200 MHz, it achieves a throughput of 38.1 Mbps, which is 22\% higher than the highest 4G bandwidth ever sampled by 4GTest \cite{huang2012mobisys}.
Meanwhile, the network stack incurs at most a few hundred microseconds delay per packet. 
Such overhead is less than 1\% of today's wireless RTT, which is from tens to hundreds of milliseconds. 

%



\subsection{Rendering}\label{sec:rendering} 

\begin{figure*}[t!]
\centering
\begin{minipage}{.45\textwidth}
        \centering
        \includegraphics[width=1\textwidth]{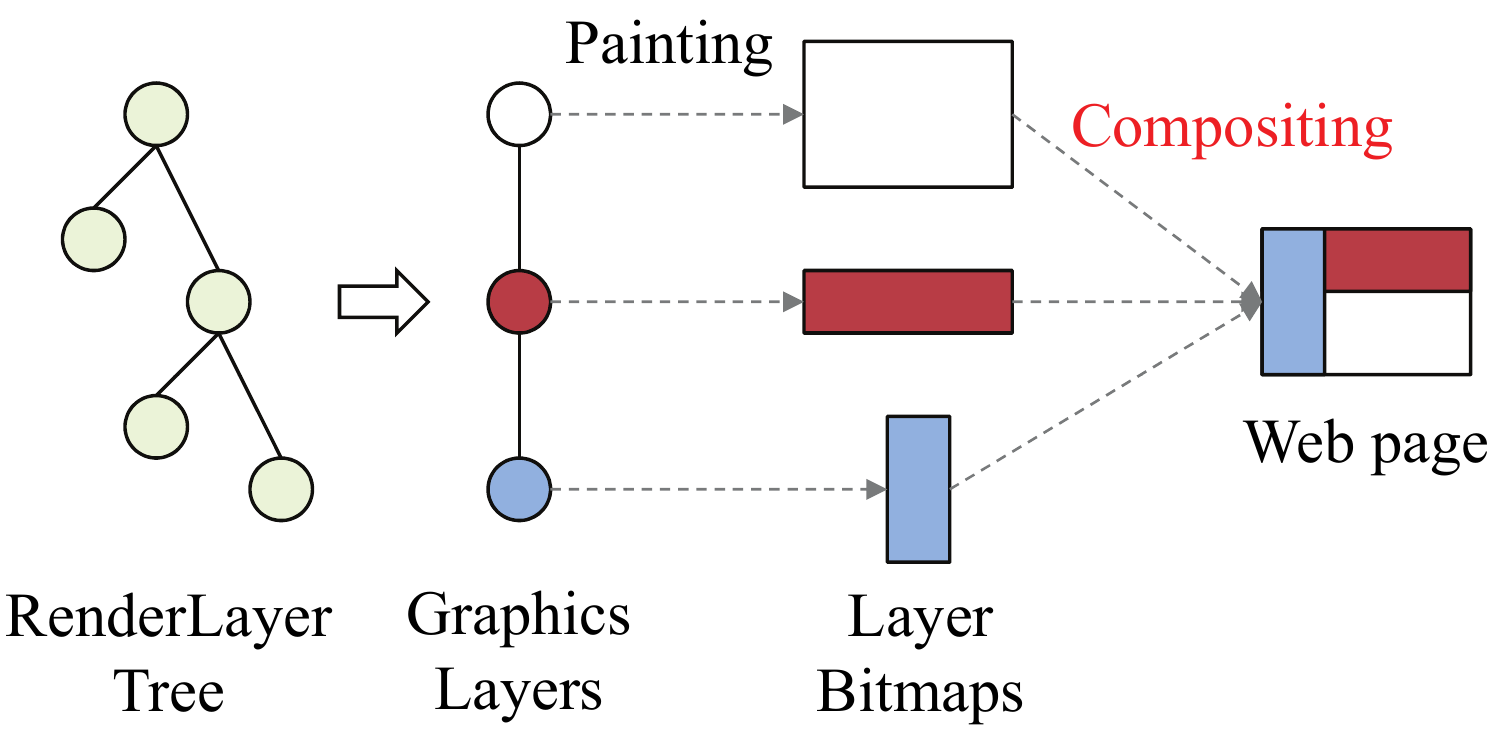}
        \caption{Hardware accelerated browser rendering. The actual hardware acceleration happens in the compositing stage.}
        \label{fig:graphics2}
\end{minipage}
\quad\quad\quad
\begin{minipage}{.45\textwidth}
        \centering
        \includegraphics[width=1\textwidth]{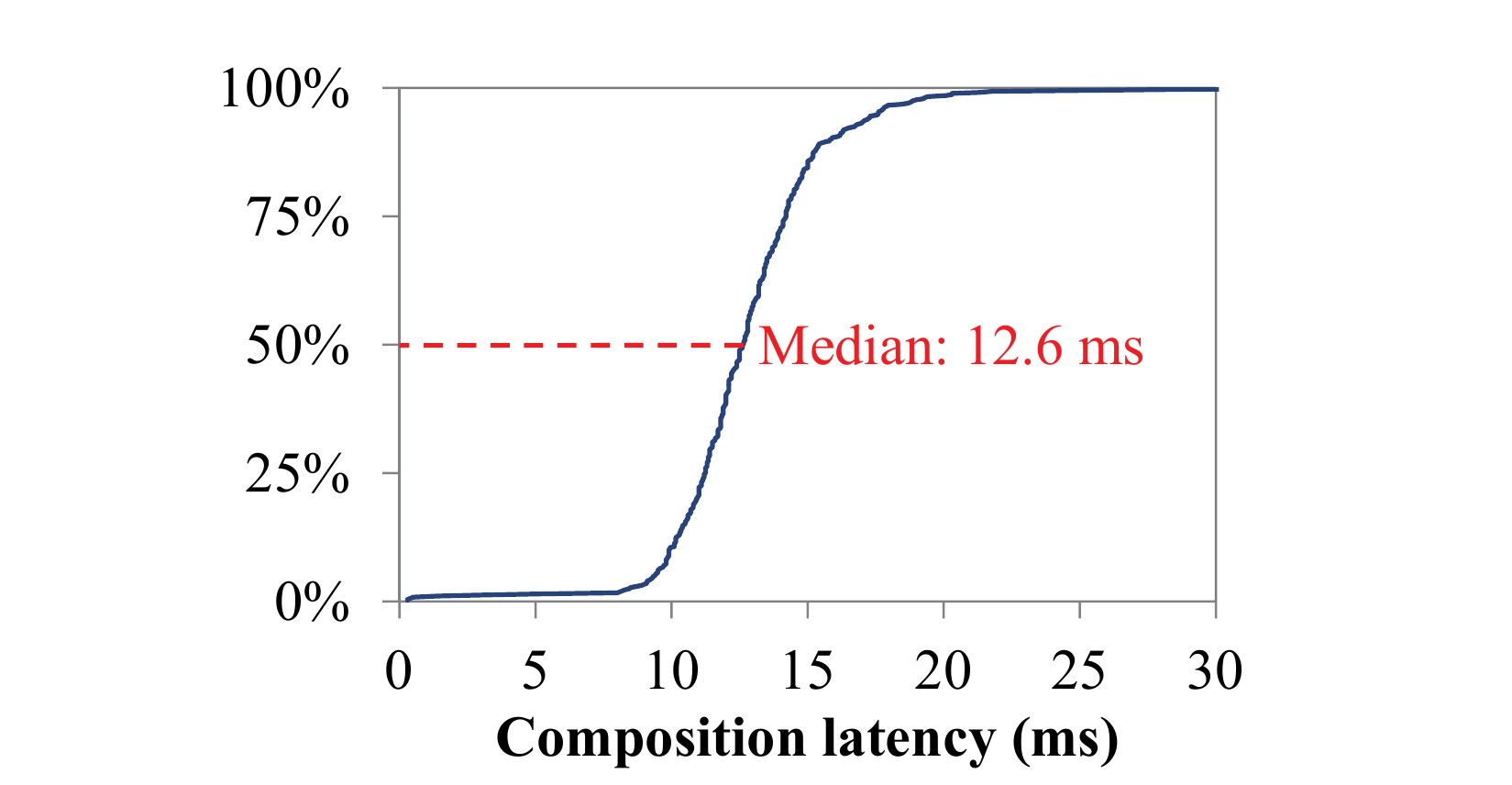}
        \caption{The CDF of the composition latency of the 2D accelerator when opening the Alexa top 500 sites' homepages}
        \label{fig:composition_latency_cdf}
\end{minipage}
\end{figure*}

Rendering is hardware accelerated by default~\cite{chromiumgpu}, as illustrated in Figure~\ref{fig:graphics2}.
The browser first creates several graphics layers from the RenderLayer tree.
Then two stages are involved: painting and compositing. 
In the painting stage, the browser paints each graphics layer into its own bitmap. 
For example, as shown Figure~\ref{fig:graphics2}, three graphics layers are painted into three layer bitmaps. 
After all layers' bitmaps are painted, in the compositing stage, the browser composites the bitmaps into one final bitmap, which is the web page.

The actual hardware acceleration happens in the compositing stage. 
The browser paints layer bitmaps by using CPU.
Then it asks GPU to composite the bitmaps into one web page. 
In theory, painting stage can also be hardware accelerated, but it is very hard to map software painting commands to commands that can be understood by GPU, e.g., OpenGL commands, and the work is still on going in industry~\cite{chromiumgpu}.

Current GPU accelerated composition only uses the 3D accelerator.
But recently, the 2D accelerator is introduced to mobile SoCs~\cite{omap4470}, which can render certain web pages with much lower power consumption, while freeing the precious resource of the 3D accelerator.
Therefore, browser rendering can exploit the 2D and 3D accelerators for better resource utilization and higher energy efficiency.

Our study of the Alexa top 500 web sites' homepages~\cite{alexa} shows that most of them can be rendered by the 2D accelerator.
In the study, we examine their rendering requirements by looking for the keywords listed in Table~\ref{tab:rendering}, which correspond to functions that are only provided by the 3D accelerator, but not by the 2D accelerator. 
As a result, out of all the 500 homepages, 449 (89.8\%) can be rendered by solely using the 2D accelerator.
In the rest of the paper, we will refer to them as 2D web pages and we will use 3D web pages for the other 51 homepages.

We also study the composition latency of the 2D accelerator.
Figure~\ref{fig:composition_latency_cdf} shows the cumulative distribution function (CDF) of the composition latency of the 2D accelerator for the Alexa top 500 web sites' homepages.
It takes the 2D accelerator 12.6 ms to composite a web page in median, 
which results in a frame rate of 30 frames per second (fps), because the browser cannot finish all the rendering tasks within one display refresh interval, i.e. 16.7 ms for 60 Hz display refresh rate.
However, the end user should still have the same smooth browsering experience because browsing 2D web pages does not require a very high frame rate and 30 fps is good enough, e.g., most motion pictures are filmed at 24 fps or 30 fps~\cite{framerate}. 
As for web pages that need a high frame rate, e.g., web gaming, they usually also contain CSS transformation, Canvas, and WebGL for animation.
In such case, the browser have switched to use the 3D accelerator to fulfill their 3D rendering requirements.

\begin{table}[t!]
	\centering
	\caption{Common HTML tags, CSS properties and JavaScript APIs relying on functions that are only provided by the 3D accelerator}
    \begin{tabular}{l|l}
		\hline 
		Categories		& Keywords	\\
		\hline\hline
		HTML Tag 		&  \texttt{Canvas~Video~Object~Embed}\\
		CSS Property 	&  \texttt{Animation~Transform~Perspective}\\
		JavaScript API	&  \texttt{WebGL}\\				
		\hline 								
    \end{tabular}
\label{tab:rendering}
\end{table}

\section{Guadalupe Design}\label{sec:design}
Exploiting hardware heterogeneity for web applications has a large design space. 
The design can be in any of the three layers: the web application, the browser or the OS.
And they can choose any heterogeneous hardware processing unit freely.
We identify four design principles that narrow down the design space, with decreasing granularities:

\begin{enumerate}
\item Make heterogeneity transparent to web developers.
\item Let the browser manage heterogeneous hardware.
\item Determine the mapping pod statically.
\item Choose hardware capability at run time.
\end{enumerate}

After discussing the principles and how we derive our design, Guadalupe, based on those principles, we describe the prototype browser implementation of the design in Section~\ref{sec:implementation}.


\begin{figure*}[t!]
\centering 
\includegraphics[width=0.9\textwidth]{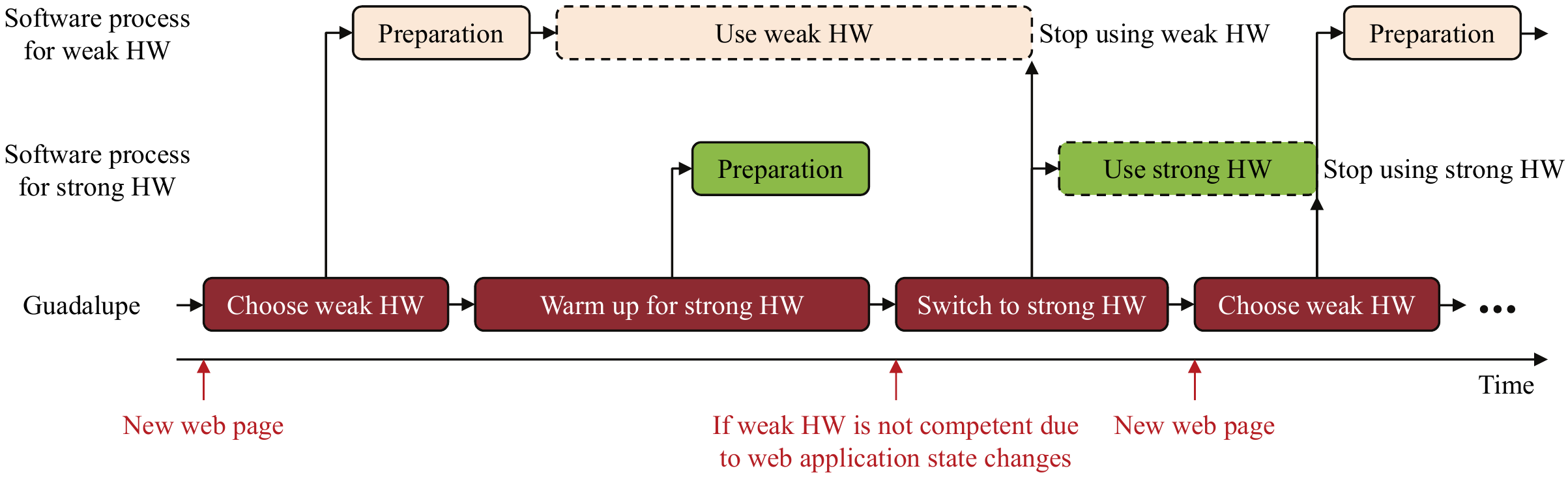}
\caption{Guadalupe design for each mapping pod. All three software processes are running on the CPU, except the function blocks with dashed border, which run on the corresponding strong or weak hardware.}
\label{fig:guadalupe_design}
\end{figure*}

\subsection{Make heterogeneity transparent to web\\ developers}\label{sec:transparent}
We seek to free web applications from the management of heterogeneous hardware, in terms of both policy, e.g. decide which hardware to use, and mechanism, e.g. switching among hardware during execution.
Our top rationale is to ease web application development: 
in face of today's fast-evolving, diverse mobile platforms,
it is virtually impossible for web developers to foresee users' platforms, 
let alone optimize applications for platform-specific hardware heterogeneity.
This rationale is also consistent with a key goal of HTML5, namely a clean separation between web application code and lower-level, platform-specific mechanisms.

Besides, current web applications can dynamically change the web content and behavior in response to the user interactions. 
The policy and mechanism should be able to provide transparent dynamic utilization of the heterogeneous hardware processing units in case of web application state change and require no effort from the web developers.

\subsection{Let the browser manage heterogeneous\\ hardware}
Given that hardware heterogeneity are made transparent to web applications, we further argue that the browser, rather than the OS, should directly manage the heterogeneity, as will be discussed below.

\textbf{Policy}.
The browser should always impose the \textit{policy}, i.e., choosing the most suitable hardware for the given mapping pod, because web application information is critical in making the policy.
This information includes performance hints, 
e.g., application behavior and future resource demands, 
as well as the interpretation of application internal state, e.g., its data structures.
Compared to the underlying OS, the browser 1) is much closer to web applications as they run in the same address space, thus having better insight into the web application, and 2) is equipped with web-specific knowledge. 
Taking resource loading as an example, with the information of resource dependency, the browser is able to infer dependencies among loading requests and thus predicts CPU utilization during loading. 


\textbf{Mechanism}.
In many cases, the browser must also implement the \textit{mechanism}, 
including translating application code to heterogeneous hardware primitives, 
supporting switch among hardware processing units during execution, etc.
As hardware are increasingly specialized for applications, such mechanisms are more likely to require deeper application knowledge, 
which is even less likely available to low-level, general-purpose system software such as the OS.
For example, the appearance of web applications are encoded in tree structures, which have to be translated for the intended graphics accelerator for rendering.
Those tree structures are complicated 
and thus can hardly be pushed down to general-purpose OSes.

%
%
We see a supportive evidence of this principle: existing OSes choose not to provide unified abstractions for any hardware specialization, except for general-purpose processors; rather, existing OSes treat them as separated I/O devices behind individual driver interfaces.
We believe one root reason is difficulties in 
taking the application knowledge into the OS.

Our direct hardware management principle is also an application of the well-known end-to-end argument~\cite{saltzer1984tocs}. 
In our case, higher-level web software layers have richer knowledge on exploiting hardware heterogeneity. 
To leverage such knowledge, we push the responsibility of heterogeneity management upwards in the software stack, so that it is close to, but not into, web applications.


\subsection{Determine the mapping pod statically}
To utilize hardware heterogeneity, a \textit{mapping pod} needs to be well defined.
A mapping pod is a set of browser functions that can be mapped onto a hardware specialization mentioned in Section~\ref{sec:background_hetero}.
The boundary of a mapping pod is determined by the natural boundary imposed by the hardware, but the spectrum of the boundary choices is very wide.
On one extreme, the boundary can be chosen at the process level.
The browser process takes the URL and shows the web page.
However, it can only be mapped to the general purpose CPUs and cannot take advantage of the specialized hardware processing units available.
On the other extreme, the boundary can be chosen at the instruction level.
Each instruction can be mapped to different hardware.
However, the strong dependency among different instructions could lead to huge overhead from choosing and switching among hardware units.

We argue that the browser designer should define the mapping pod and map it to the appropriate hardware specialization statically.
Browser designers understand the browser functionalities and hardware primitives very well.
During the design time, they can find the best mapping of a mapping pod and a hardware specialization to maximize performance and power efficiency.
In contrast, it is too hard for the browser to figure out the correct mapping automatically.

There are two mapping pods among the six basic browser services shown in Figure~\ref{fig:browser}.
Resource loading is a mapping pod that can be mapped to the asymmetric processors.
Rendering is a mapping pod that can be mapped to the graphics accelerators.
The rest four browser services cannot benefit from any of the current available heterogeneous hardware processing units, yet.
So they belong to no mapping pod.
The two add-on services, video decoding and image processing are also mapping pods and they can be mapped to the DSPs.
In case a new hardware specialization emerges and can benefit other browser functionalities, a new mapping pod can be created by the browser designer to utilize those heterogeneous hardware units.

\subsection{Choose hardware capability at run time}
After determining the mapping pod and map it to a hardware specialization statically, the browser should choose hardware capability for the mapping pod at run time.
As discussed in Section~\ref{sec:background_hetero}, a hardware specialization includes multiple processing units with different capabilities.
The strong one has more functionalities or better performance, but consumes more energy. 
Based on the run-time web application state, the browser can make the best choice of the hardware capability for better performance and power efficiency.

We next use the two cases studied in Section~\ref{sec:case} to exemplify the principle.
Based on the typical workload pattern a browser has, the browser can load the main resource with the weak processor for power efficiency and load the subresources with the strong processor for good performance.
As for rendering, the browser can render 2D and 3D web pages with the 2D and 3D accelerators, respectively.
In case any 3D rendering requirement is added to a 2D web page, e.g., by user interaction or animation, the browser can detect the change of the application state and switch to use the 3D accelerator on demand.
In this case, the browser not only makes the 3D accelerator available for other applications, but also improves energy efficiency.

\subsection{Applying the principles}
Guided by the four principles discussed above, we have designed Guadalupe to utilize the hardware candidate with desired capability based on the run-time dynamics of the web application state for better resource utilization and energy proportionality.

Guadalupe always starts from the weak hardware for each web page, and switches to the strong one on demand.
The rationale is that we exploit hardware heterogeneity in the browser to make more computing resources available and improve energy proportionality. 
Choosing the weak hardware from the beginning for each web page is more energy efficient and frees the strong hardware for other services. 
In case that the weak hardware cannot provide the desired performance or cannot fulfill certain functionalities due to its limited capability, the browser will switch to use the strong hardware.
Once switching to use the strong hardware, Guadalupe will not switch back to the weak one until a new web page is open, because the browser service's requirement for the current web page will not be reduced.

Figure~\ref{fig:guadalupe_design} illustrates the Guadalupe design for each mapping pod.
When starting to open a web page, Guadalupe chooses the weak hardware for the mapping pod.
Guadalupe monitors the web application state and checks whether the weak hardware is competent.
In case the weak hardware cannot provide the desired performance or functionality, Guadalupe switches to use the strong hardware on demand and all the data structures needed by the strong hardware will be prepared.
When a new web page is open, Guadalupe switches back to use the weak hardware and previously used data structures are cleared.

The key challenge in Guadalupe design is efficient switch.
While getting better resource utilization and energy proportionality, Guadalupe should also switch from the weak hardware to the strong one with low overhead.

The switching overhead mainly comes from the data structure preparation for the strong hardware.
We optimize Guadalupe by redundantly preparing data structures for the strong hardware, before the switch happens.
So when the switch takes place, it incurs low overhead and provides a smooth transition between the two hardware units, which may not even be noticed by the user.

It is tempting to redundantly prepare the data structures for the strong hardware right before the switch is needed.
However, each web application has different state, and user interaction with the web also changes the web application state dynamically, making it impossible for the browser to predict exactly when the switch is going to happen.
Therefore, we design Guadalupe to prepares the required data structures for the weak and strong hardware processing units simultaneously.
Those redundantly prepared data structures may not be used by the strong hardware.
But in case the switching is needed, the data structures needed by the strong hardware is guaranteed to be ready, leading to low switching overhead.

Redundant preparation ensures the transparent and smooth switching. 
But it also incurs several overheads. 
The browser needs more CPU power and memory to prepare and store the redundant data structures. 
For the resource loading case, the redundantly prepared data structures are the URLs, which are simple and small. 
For the rendering case, Guadalupe needs to prepare the layer bitmaps for the 3D accelerator.
We evaluate redundant preparation in Section~\ref{sec:evaluation} and show that the switch is fast and the overhead is small.

\section{Implementation}\label{sec:implementation}

We next discuss a prototype implementation of Guadalupe design: Guadalupe browser. 
We use TI Blaze Tablet~\cite{tiblaze} with OMAP4470~\cite{omap4470} as the mobile device. 
OMAP4470 features two types of asymmetric processors: dual Cortex-A9 and dual Cortex-M3 processors.
It also has a 3D accelerator based on PowerVR SGX544 core from Imagination Technologies and a 2D accelerator based on GC320 2D core from Vivante Corporation. 
One can use OpenGL API to use the 3D accelerator and use BLTsville API~\cite{tibltsville} to use the 2D accelerator. 

The Guadalupe browser implementation is based on Chrome for Android beta~\cite{chromeandroid}, which runs in Android ICS on the Blaze tablet. 
Chrome for Android is not fully open sourced yet~\cite{chromemobile}, especially the Java side code. 
We are able to pull a snapshot of the Chrome for Android beta source code~\cite{beverloo2012chrome}. 
Combined with some other Chromium source code, we manage to modify and compile its C++ side source code and produce the shared library \texttt{libchromeview.so}. 
Then we push the shared library into the tablet to turn Chrome for Android beta to Guadalupe browser.

We first give an overview of the system architecture of Guadalupe browser.
Then we discuss the implementation details of Guadalupe browser for resource loading and rendering.

\subsection{Overview of system architecture}\label{sec:architecture}

\begin{figure}[t!]
\centering 
\includegraphics[width=0.45\textwidth]{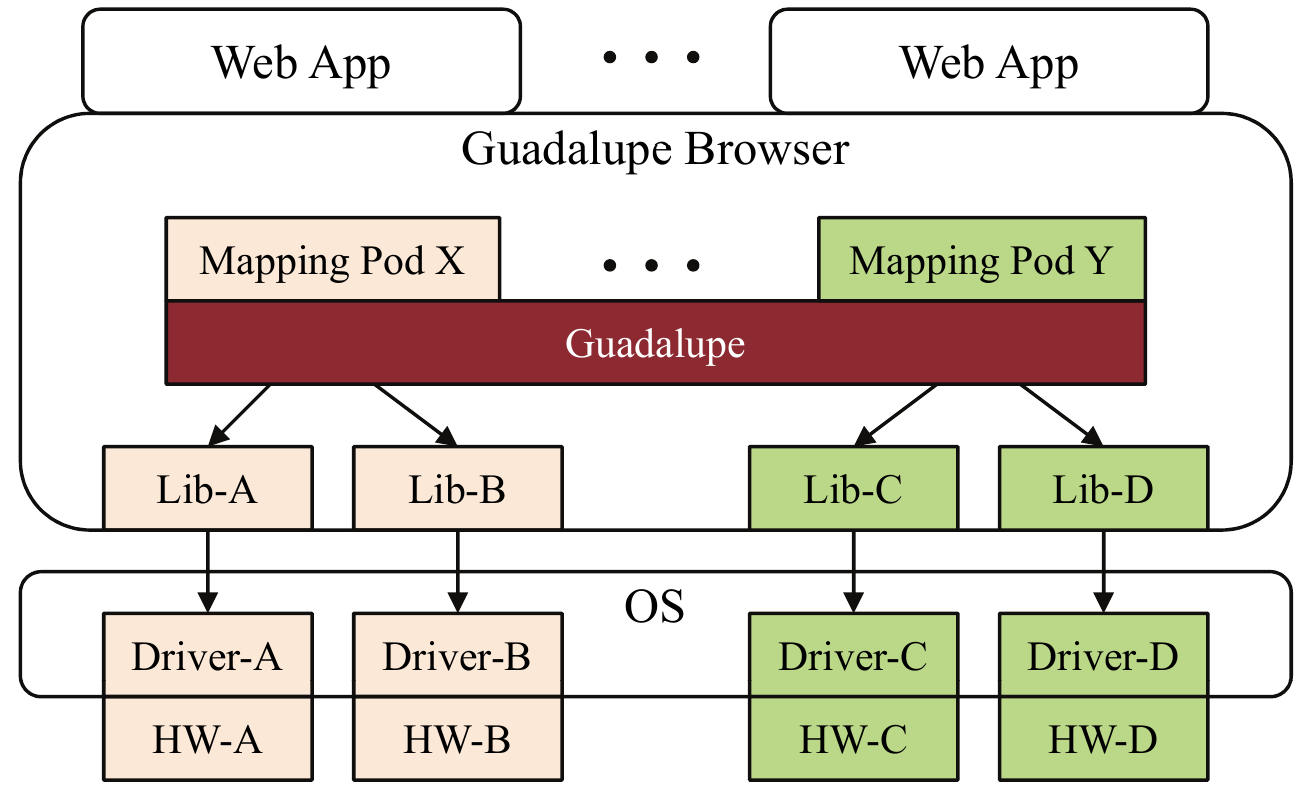}
\caption{Guadalupe system architecture. The hardware can be either CPU or I/O devices such as graphics accelerators. Correspondingly, the driver is OS service for the CPU or device driver for I/O devices.}
\label{fig:architecture}
\end{figure}

Figure~\ref{fig:architecture} illustrates the system architecture of the Guadalupe browser. 
By adding a Guadalupe layer between the mapping pod and the hardware interfaces, Guadalupe browser enables mapping pod X and Y to utilize different hardware processing units within their mapped hardware specialization.
Guadalupe layer monitors the web application state, chooses the suitable hardware for the mapping pod and switches among the hardware candidates on demand.
If multiple web applications are running on the browser, Guadalupe layer manages the hardware for them separately.

The hardware candidates can be CPU or specialized processing units like graphics accelerators, treated as I/O devices by the system.
The hardware interface between the browser and CPU is the OS itself. 
The OS may also provide further abstraction to utilize heterogeneous multi-processors in the future, e.g., asymmetric processor detection and heterogeneous architecture aware system calls.
The hardware interface between the browser and an I/O device is the device driver.
Guadalupe browser dynamically loads the corresponding hardware library into the browser's address space and uses it as the interface to talk to the device driver in the OS.
If library loading is failed, Guadalupe browser will assume that the corresponding hardware is not available on the device.

\subsection{Resource loading}\label{sec:impl_loading}
Guadalupe browser exploits asymmetric processors for resource loading by loading the initial a few resources with the weak processor and later switch to the strong processor for  subsequent resources.

In a legacy browser,   
resource loading invokes various network services, e.g., setting up TCP connections, transmitting and receiving packets, etc. 
After getting the URLs of the resources, the browser sends all the resource requests to a HTTP library, which is implemented by Chrome for Android beta.
The HTTP library keeps track of all the pending resource requests and in turn invokes the transport layer services provided by the OS through system calls.

In Guadalupe browser, the HTTP library is modified for our resource loading.
When Guadalupe browser sends the URL request to the HTTP library, the resource information is embedded in the request to indicate which processor the request should be made with.
In turn, our modified HTTP library creates and uses TCP connections on the weak or strong processor for loading the resource.

Guadalupe browser provides the policy to select processors for resource loading; 
it relies on a heterogeneity-aware OS to provide the mechanism that creates and maintains TCP connections on asymmetric processors. Such an OS, while missing as of now, is a key goal of our on-going efforts ~\cite{lin2012hotpower}.

\subsection{Rendering}\label{sec:impl_rendering}
Guadalupe browser for rendering exploits the 2D and 3D accelerators on the Blaze Tablet.
We first give a brief background of rendering on Android system.
Then we discuss the implementation details of Guadalupe browser for rendering.

\begin{figure}[t!]
\centering 
\includegraphics[width=0.45\textwidth]{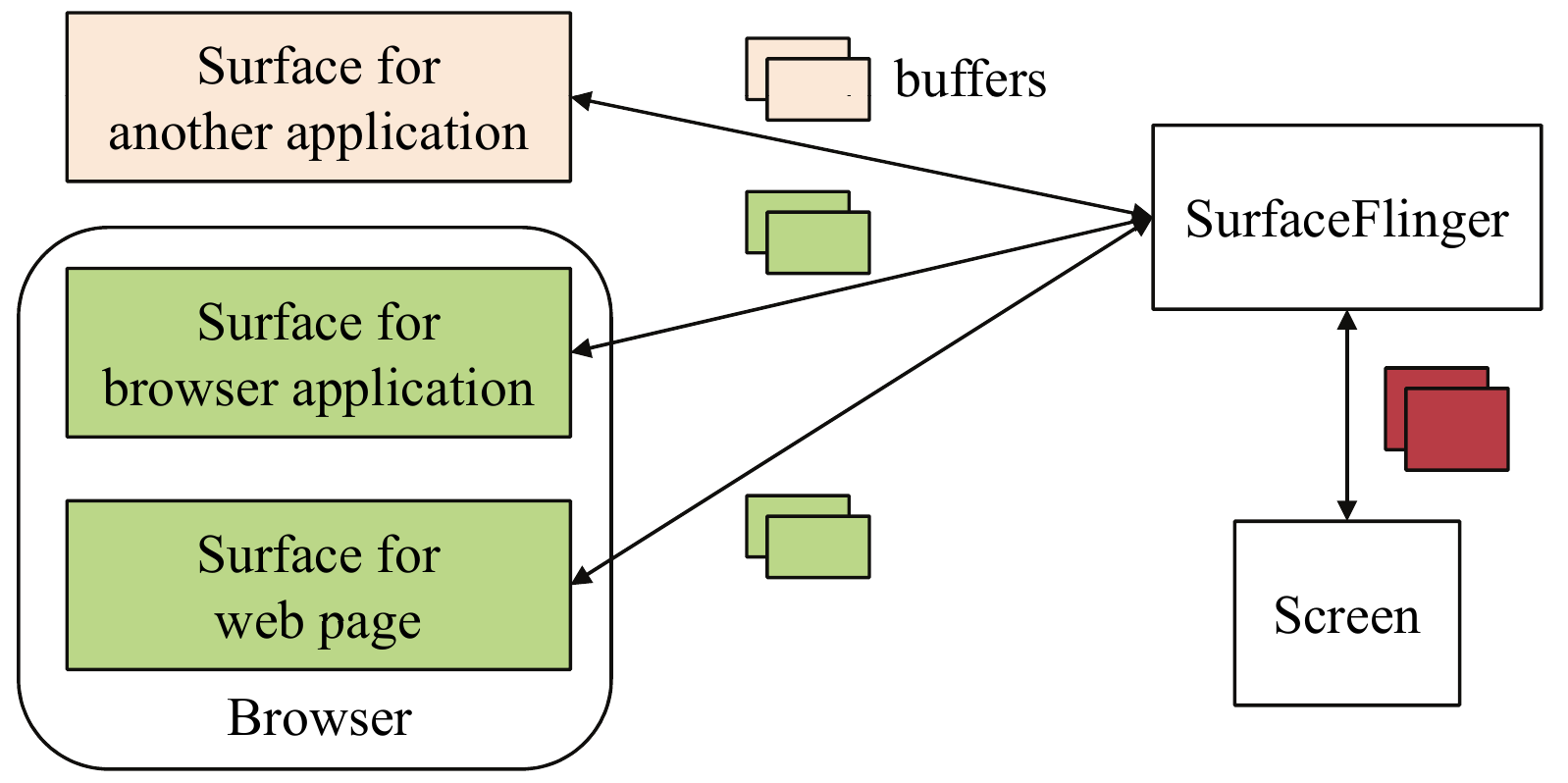}
\caption{The Android rendering framework. Browser application rendering and web page rendering are separated.}
\label{fig:androidrendering}
\end{figure}

On Android, apart from browser rendering for web pages, the browser also needs to render the application itself, i.e., the address bar, back and forward buttons, etc. 
And browser application rendering and web page rendering are separated. 
Figure~\ref{fig:androidrendering} illustrates how Android rendering works with the browser.
The center of the rendering system is SurfaceFlinger, which manages the buffers for the window system. 
The browser interacts with SurfaceFlinger through surface. First, the browser connect its surfaces to SurfaceFlinger.
Then for each frame, the browser requests application buffers from SurfaceFlinger through the connected surfaces.
After rendering the web page and the application, the browser posts the buffers to SurfaceFlinger. 
SurfaceFlinger takes the buffers from different applications, composite them into one frame buffer and posts it onto the screen.

When the browser is started, Guadalupe loads both BLTsville library~\cite{tibltsville} and OpenGL library~\cite{OpenGL} into its address space, and places separate switching hooks for the 2D and 3D accelerators.
A switching hook is basically a callback, which is invoked for later potential switch on demand.
When Guadalupe browser opens a web page, it allocates the web page's surface to the 2D accelerator for rendering. 
During the page opening, Guadalupe layer prepares the layer bitmaps for both 2D and 3D accelerators. 
Guadalupe monitors the changes of the web page state.
Once 3D rendering requirement is detected, Guadalupe browser switches to use the 3D accelerator by disconnecting the web page surface's current connection to SurfaceFlinger, reallocating the surface to the 3D accelerator, and connecting the reallocated surface to SurfaceFlinger again.
Note that Guadalupe manages the 2D and 3D accelerators for web page composition only, which is separated from browser application rendering.

\section{Evaluation}\label{sec:evaluation}
We evaluate the two key aspects of Guadalupe browser, rendering and resource loading.
To evaluate rendering, we run Guadalupe browser on an OMAP4-based TI Blaze Tablet~\cite{tiblaze} 
with Android ICS.
Our results show that compared to legacy browsers, Guadalupe browser provides comparable performance and better resource utilization, while incurring lower power consumption and little overhead.
For evaluating resource loading, we employ a combination of estimation and micro-benchmarks on OMAP4 to show the significant benefits from Guadalupe, despite we do not yet have a complete resource loading implementation.

\subsection{Rendering}
Guadalupe browser for rendering is evaluated in four aspects: performance, resource utilization, efficiency and overhead. 
To compare with Guadalupe browser, we use Chrome for Android beta~\cite{chromeandroid} and we will use mobile Chrome to refer to it in the rest of the paper.
We instrument the browsers to measure the page load time, the composition latency of the 2D accelerator and overhead.
We use PVRTune~\cite{pvrtune} to monitor the 3D accelerator activities and OMAPCONF~\cite{omapconf} for bandwidth consumption of the accelerators.
During the experiments, the tablet is connected to the local Ethernet network though WiFi interface. 
We have set up a local web page replay server~\cite{webreplay} to remove the network variations. 
We first record the Alexa top 500 web sites' homepages~\cite{alexa} with the web page replay server. 
Then we configure the tablet to use the web page replay server as the proxy, which always serves the resource requests from the tablet with pre-recorded resource files. 
We have run though the Alexa top 500 web sites' homepages for 5 rounds for the evaluations.

While providing the same performance as mobile Chrome, Guadalupe browser makes more hardware resources available to other applications, is more power efficient, and incurs little overhead.

\subsubsection{Performance}
\begin{figure}[t!]
\centering 
\includegraphics[width=0.45\textwidth]{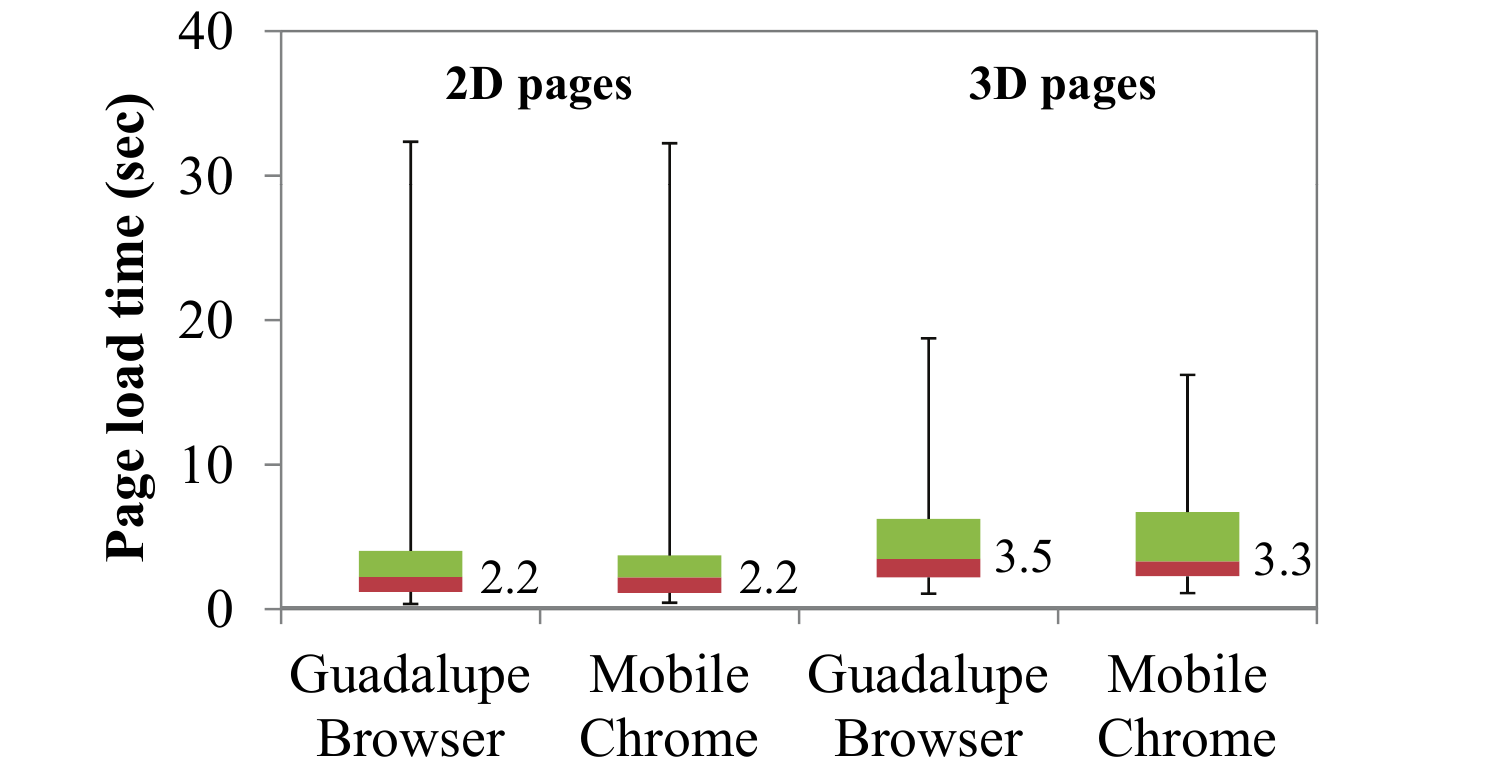}
\caption{The web page load time of the Alexa top 500 web sites' homepages, with data points at min, 25th percentile, median, 75th percentile and max values. The median values are labeled.}
\label{fig:page_load_time}
\end{figure}

Guadalupe browser performs as good as mobile Chrome in terms of web page load time, as shown in Figure~\ref{fig:page_load_time}. 
For 3D web pages, Guadalupe browser starts with the 2D accelerator and switches to use the 3D accelerator after detecting 3D rendering requirements from the web application state.

With 30 fps frame rate, Guadalupe browser also provides the same smooth browsing experience, because 2D web pages does not require a very high frame rate, as discussed in Section~\ref{sec:rendering}.

\subsubsection{Resource utilization}\label{sec:resource_utilization}
\begin{figure}[t!]
\centering 
\includegraphics[width=0.45\textwidth]{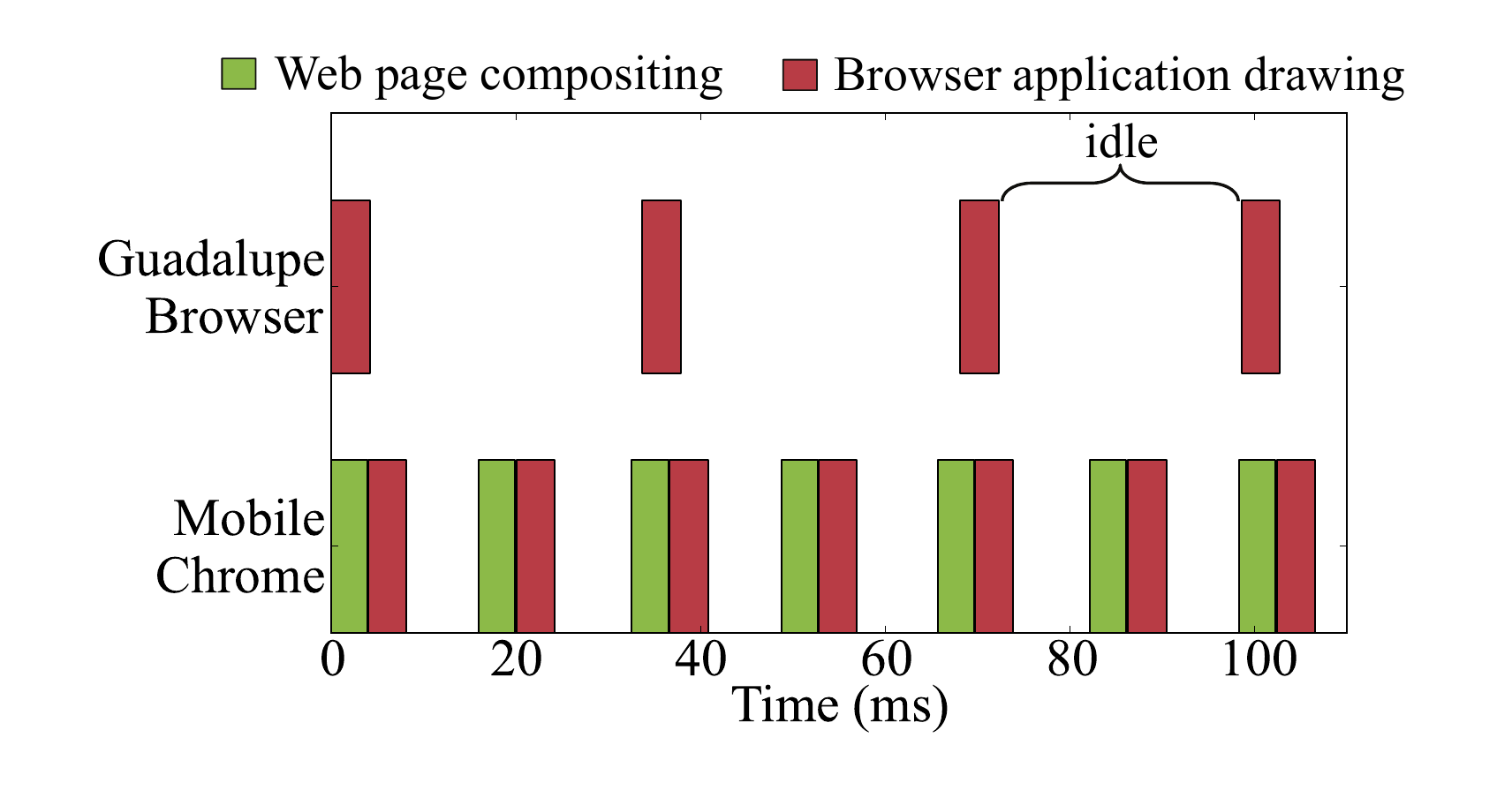}
\caption{The 3D accelerator activities in Guadalupe browser and mobile Chrome.}
\label{fig:resource_utilization}
\end{figure}

Guadalupe browser makes more hardware resources available to other applications.
Figure~\ref{fig:resource_utilization} shows an example of the 3D accelerator activities when Guadalupe browser and mobile Chrome are activly compositing the web page. 
By utilizing the 2D accelerator, Guadalupe browser reduces the usage of the 3D accelerator by 75\% and frees it for potential 3D tasks from other applications.
One reason is that Guadalupe browser involves only one 3D accelerator activity for each frame, i.e., drawing the browser application.
The web page composition is done by the 2D accelerator.
In contrast, mobile Chrome involves two 3D accelerator activities for both browser application drawing and web page composition.
The other reason is that Guadalupe browser has smaller frame rate, as discussed in Section~\ref{sec:rendering}.
But even if the two browsers has the same frame rate, Guadalupe browser still reduces the 3D accelerator usage by at least 50\%.

Due to the limitations of current Android system, only the front application can be actively rendered. All background applications are paused and will not be rendered, making the 3D accelerator not fully utilized.
However, as future mobile devices start to support split screen and external monitor for the second front application, the idle time of the 3D accelerator freed from Guadalupe browser can be better utilized.
For example, both Samsung~\cite{samsungnote} and Microsoft~\cite{microsoftsurface} start to sell tablets with split screen capability. 
While browsing a web page by using the 2D accelerator on one side of the tablet, the user can play a video, a 3D game or any 3D application on the other side of the tablet simultaneously, without too much contention for the 3D accelerator. 

We estimate what frame rate the other 3D application can achieve.
We use a 3D cube rotation application as the benchmark, which achieves 60 fps on Blaze Tablet.
For each frame, the 3D accelerator spends 10 ms to draw the application and 6.7 ms for SurfaceFlinger composition.
We assume that the browser and the benchmark can run side by side on Android and the browser's composition need is satisfied by the 3D accelerator first.
With Guadalupe browser (30 fps) running on the other side, the frame rate of the benchmark is 52 fps, which is very close to its original frame rate.
However, with mobile Chrome (60 fps), the frame rate of the benchmark drops to 6 fps.
Even if we set mobile Chrome's frame rate to 30 fps, the same as Guadalupe browser, the frame rate of the benchmark still drops to 44 fps.
Therefor, Guadalupe browser increases the frame rate of the other 3D application by 18\% to 767\%.

%

\begin{figure*}[th]
\centering
\subfigure[Guadalupe Browser]{
	\centering 
	\includegraphics[width=0.45\textwidth]{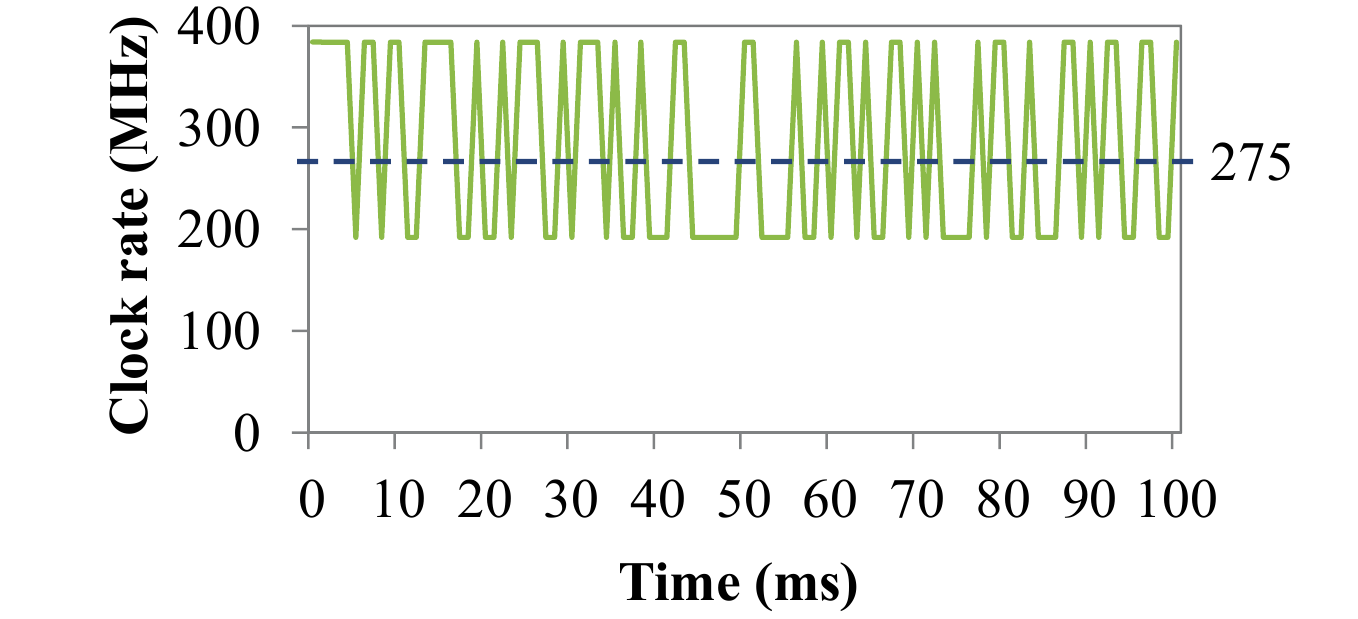}
    \label{fig:gpu_frequency_guadalupe}
}
\hspace{6mm}
\subfigure[Mobile Chrome]{
	\centering 
	\includegraphics[width=0.45\textwidth]{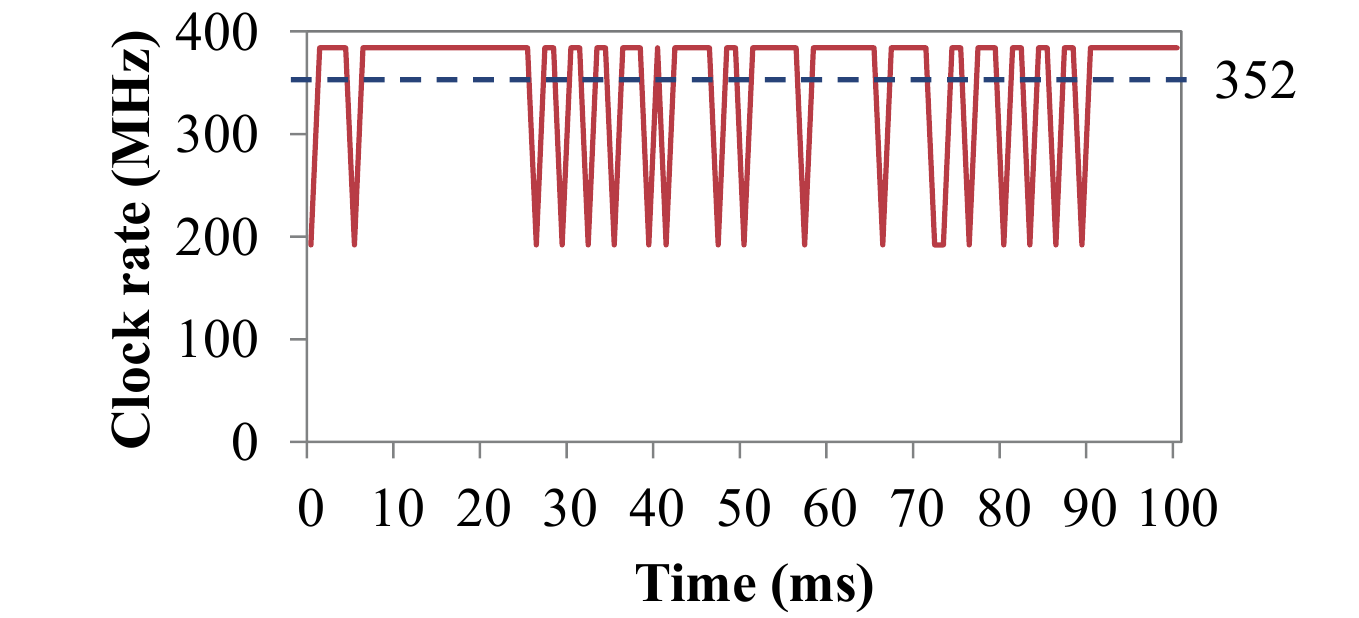}
    \label{fig:gpu_frequency_chrome}
}
\caption{The clock rate of the 3D accelerator when using \subref{fig:gpu_frequency_guadalupe} Guadalupe browser and \subref{fig:gpu_frequency_chrome} mobile Chrome. The average clock rate is labeled in the figure.}
\label{fig:gpu_frequency}
\end{figure*}

\begin{figure*}[th]
\centering
\subfigure[3D accelerator]{
	\centering 
	\includegraphics[width=0.45\textwidth]{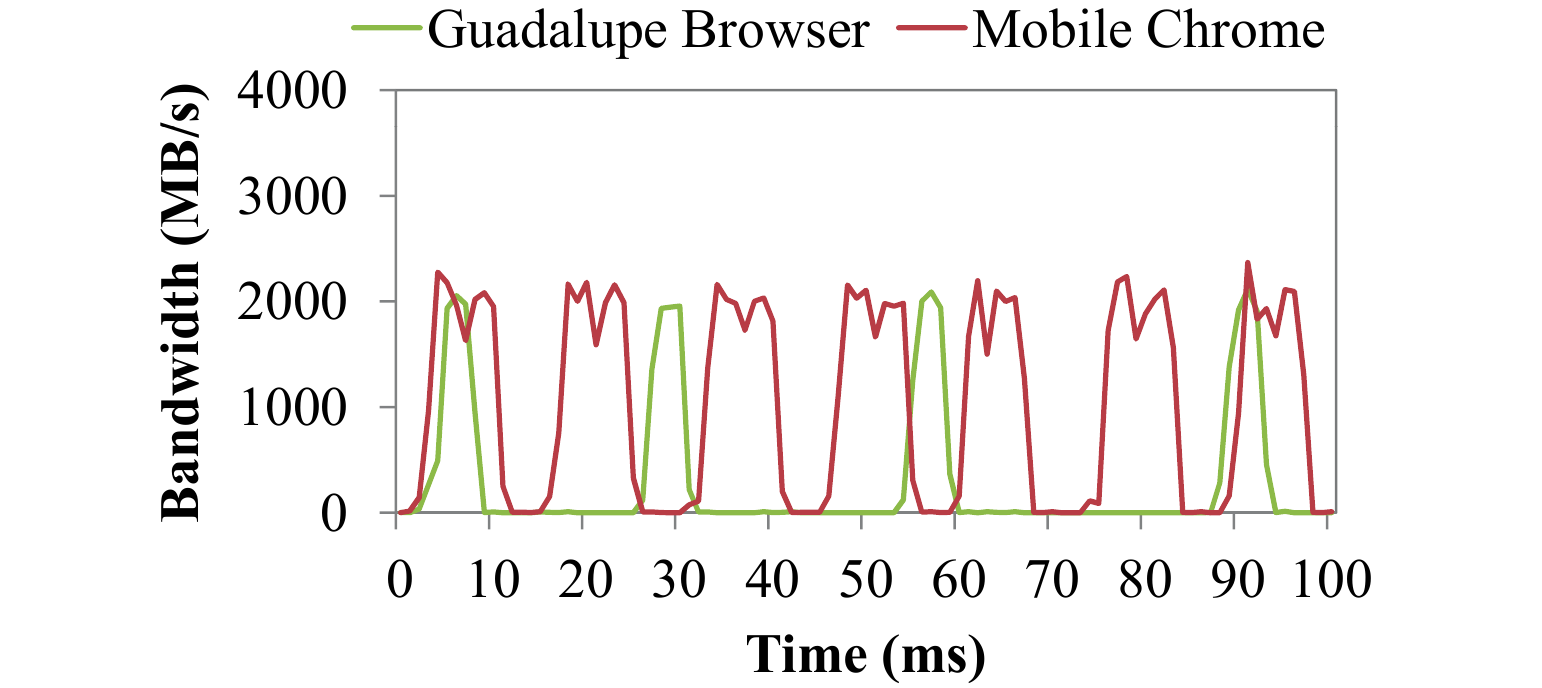}
    \label{fig:gpu_bandwidth}
}
\hspace{6mm}
\subfigure[2D accelerator]{
	\centering 
	\includegraphics[width=0.45\textwidth]{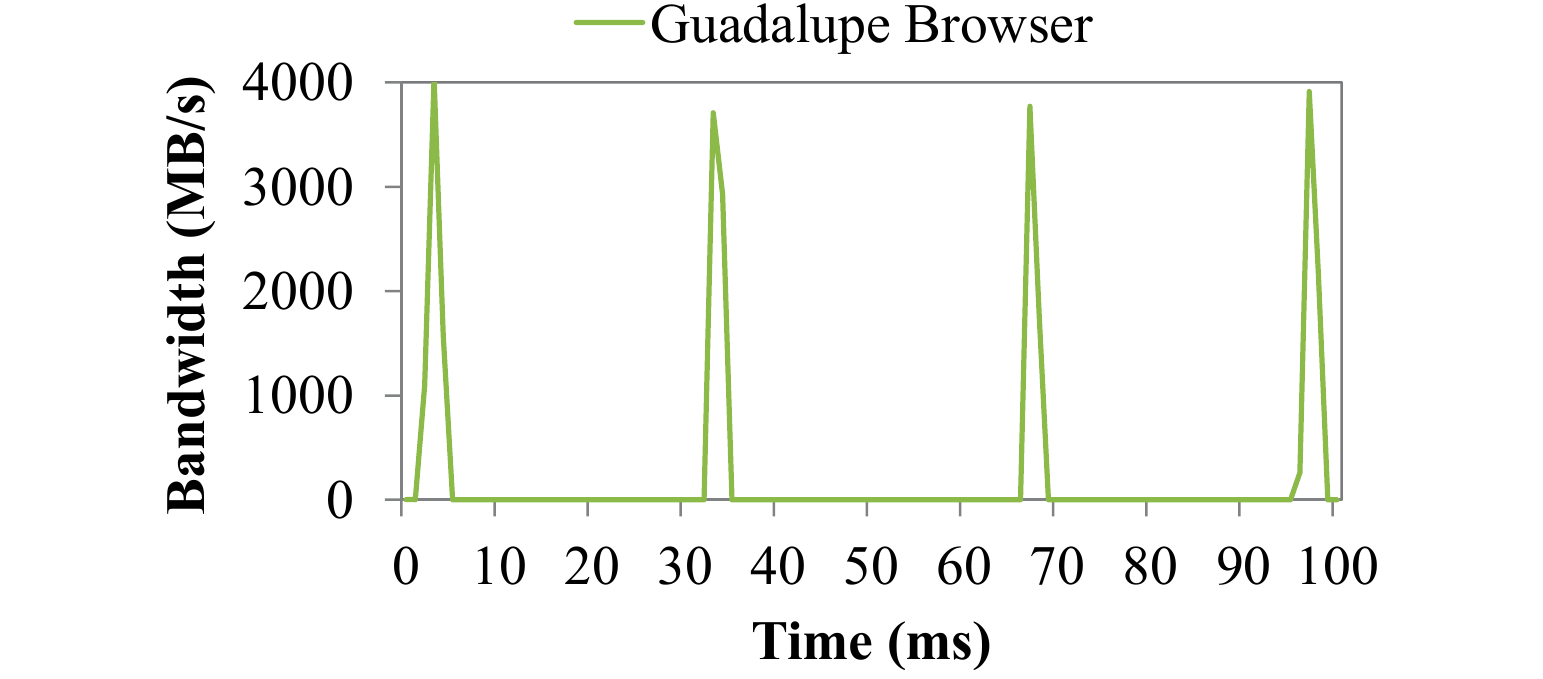}
    \label{fig:gc320_bandwidth}
}
\caption{The bandwidth consumption of \subref{fig:gpu_bandwidth} the 3D accelerator and \subref{fig:gc320_bandwidth} the 2D accelerator when using Guadalupe browser and mobile Chrome. Mobile Chrome does not use the 2D accelerator.}
\label{fig:bandwidth}
\end{figure*}

\subsubsection{Efficiency}
Guadalupe browser is more power efficient than mobile Chrome. 
As mentioned in Section~\ref{sec:resource_utilization}, Guadalupe browser still needs to use the 3D accelerator to draw the application. 
However, it can utilize the more power efficient 2D accelerator to composite web pages.

The 2D accelerator only consumes tens of mW while the 3D accelerator typically consumes several hundred mW.
On OMAP4470~\cite{omap4470}, the 3D accelerator consumes over 12 times more active power than the 2D accelerator.
Due to the confidential nature of power consumption numbers of TI chip-sets involved in this study, we could not publish exact power numbers of the accelerators.
However, we can compare the browsers' efficiency by showing the accelerators' clock rate, bandwidth consumption and estimated relative power consumption for composition.
We also estimate how much energy Guadalupe browser can save for the entire system.

We use scrolling as the benchmark to ask the accelerators to continuously composite a 2D web page. 
Before each experiment, we scroll through the web page, so that all the data structures needed by the accelerators are already prepared and any further scrolling will not generate new content. 
Then we continuously scroll the web page for over five seconds for measurements.

The 3D accelerator in Guadalupe browser runs in a much lower average clock rate and consumes much less bandwidth, as shown in Figure~\ref{fig:gpu_frequency} and Figure~\ref{fig:gpu_bandwidth}. 
Besides, the 2D accelerator in Guadalupe browser moves the web page content twice faster than the 3D accelerator in mobile Chrome, as shown in Figure~\ref{fig:bandwidth}.
Furthermore, we esimate that the 2D accelerator is 6 times more power efficient than the 3D accelerator for single frame web page composition.

We also estimate how much energy Guadalupe browser can save for the entire system.
Since TI Blaze Tablet~\cite{tiblaze} is an industry prototype, its power consumption is not representative and not optimized.
Instead, we measure the power consumption of Samsung Galaxy Nexus, whose SoC belongs to the same OMAP4~\cite{omap4} family.
Its total system power consumption is 1700 mW while browsing the web over WiFi network with full brightness of the screen.
For one second of active composition, e.g., due to scrolling or 2D animation, Guadalupe browser saves 80 mJ. 
Therefore, Guadalupe browser saves 4.7\% energy consumption of the entire mobile system.
Guadalupe browser for rendering does not save much energy for the entire mobile device because other hardware components, e.g., the display, consume most of the energy.
But more importantly, Guadalupe design creates the great opportunity for many other mapping pods to improve their energy proportionality.


\subsubsection{Overhead}\label{sec:overhead}

\begin{figure}[t!]
\centering 
\includegraphics[width=0.45\textwidth]{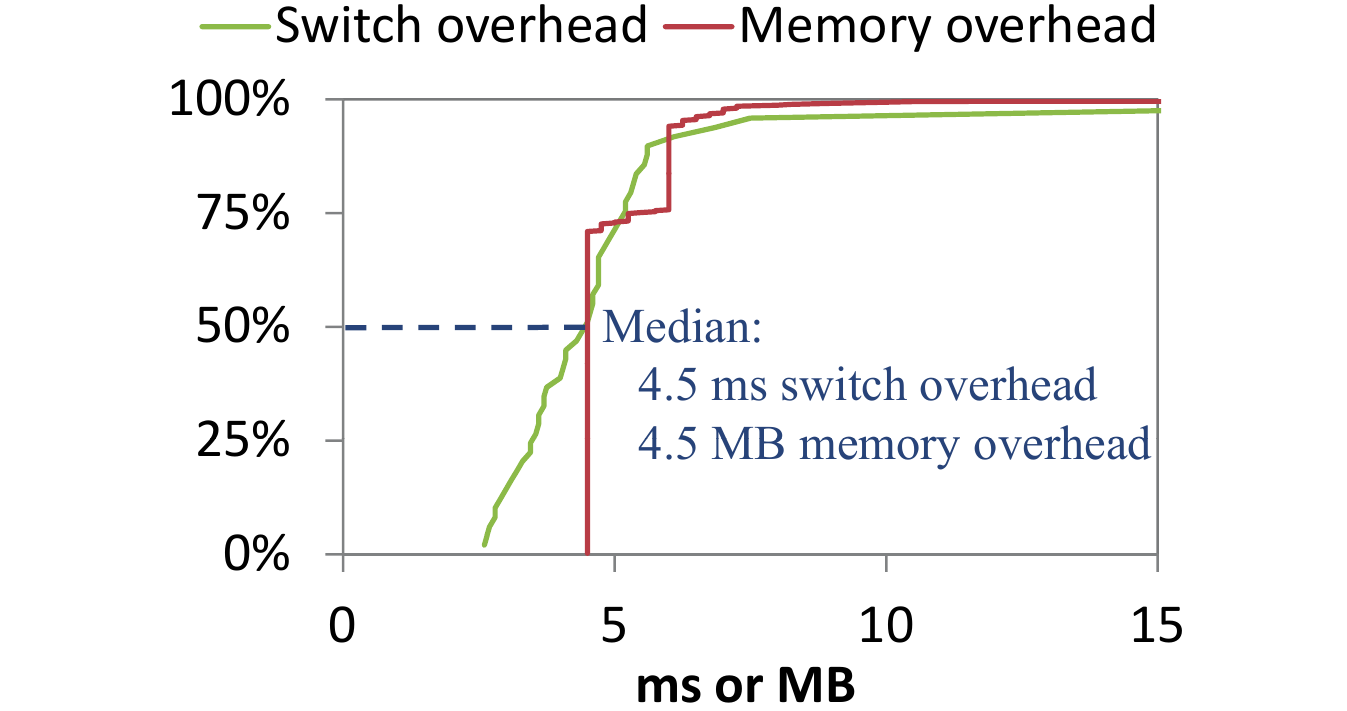}
\caption{The CDF of the overhead of Guadalupe browser when opening the Alexa top 500 sites' homepages.}
\label{fig:overhead_cdf}
\end{figure}

Guadalupe browser ensures efficient switch with little overhead.
The switch overhead is the time between when the switch need is detected and the time when the switch is finished, which is shown in Figure~\ref{fig:overhead_cdf}. 
The switch of Guadalupe browser is very fast and only takes 4.5 ms in median. 
This is negligible comparing to the web page load time.

The overhead of redundant preparation are mainly extra memory allocation due to the data structures prepared for the hardware candidates, which is also shown in Figure~\ref{fig:overhead_cdf}. 
In median, Guadalupe browser consumes 4.5 MB for each web page, which is negligible comparing to current mobile devices' memory size. 
The minimum memory overhead is also 4.5 MB because a web page needs at least one graphics layer, which consumes 4.5 MB memory.
Even if the memory overhead becomes huge due to the large number of graphics layers for the web page, Guadalupe browser can always set a memory limit and stop redundant preparation to avoid excessive memory allocation.

The CPU power consumption overhead from redundant preparation is also negligible.
The data structure preparation for the 2D accelerator and other browser services like parsing and layout consume over 50\% CPU utilization and they have already pushes the CPU to its peak clock rate.
Therefore, redundantly preparing data structures for the 3D accelerator simultaneously incurs little CPU power consumption overhead.

\subsection{Resource loading}
We evaluate resource loading through a combination of estimation and micro-benchmarks on TI OMAP4.
We does not consider the goal of making more hardware resource available, 
as we expect Cortex-A9's resource is offline to save power when Cortex-M3 performs loading.

\subsubsection{Energy efficiency}

As implied in Section \ref{sec:case}, 
Guadalupe is able to greatly improve energy proportionality~\cite{lin2012hotpower} by mapping resource loading to weak processors; 
we next estimate the energy reduction of this design, as compared to pinning resource loading on A9, i.e., the legacy case.
We make two assumptions: 
1) the same implementation of resource loading is mapped to either A9 or M3, and 
2) in considering the effect of DVFS, 
we use the corresponding clock rate and power scaling factors published by TI for the last generation OMAP SoC \cite{TIPET}.

Our estimation results are shown in Table \ref{tab:eval-resource-loading-energy-reduction}.
In deriving the results, we first estimate the power of resource loading on A9.
As resource loading is a light workload, 
we assume that A9 can perform it with the lowest clock rate at 200 MHz, i.e., its most energy efficient state.
Given that A9 running at 1 GHz typically consumes 250 mW \cite{armca9}, 
by applying 5.8X clock scaling and 11X power scaling, we estimate that
A9 running at 200 MHz consumes 22.5 mW.

\begin{table}[t!] 
	\centering
	\caption{Estimated power and energy reduction of resource loading on Cortex-M3, as compared to on Cortex-A9}	
    \begin{tabular}{l | c c c |  c}
		\hline 		    		
        & & M3 & & A9\\
        		\hline 		    		\hline 		    
Clock rate (MHz) & 	34.7	 &	100 				 & 200 	 & 200				\\
		Power (mW) & 1.7 & 7.2 & 19.0 	& 22.5 \\
		Energy reduction & 56.2\%   & 39.3\% & 15.7\% & -- \\
		\hline 						
    \end{tabular}%
\label{tab:eval-resource-loading-energy-reduction}
\end{table}

We next estimate the power of resource loading on M3.
Running at 200 MHz, M3 typically consumes 19 mW \cite{armcm3}.
Applying scaling factors from \cite{TIPET}, we estimate that M3 consumes 7.2 mW at 100 MHz and 1.7 mW at 34.7 MHz.
Comparing the power consumption of two cases and taking account into the clock differences, we conclude that mapping resource loading to M3 can achieve as high as 56.2 \% energy reduction as compared to pinning it on A9.

In practice, such an energy reduction will be even higher, 
due to processor idle periods that frequently occur in resource loading.
In such short idle periods, A9 either spends high idle power ($\sim$11 mW \cite{TIPET}) 
or frequently enters and exits deep-sleep power state, both of which are energy hungry.
In comparison, M3 has 10X less idle power ($<$ 1 mW) while being able to perform much more efficient power state transitions, thanks to its lightweight architecture.

We stress that the above estimation must be based on mapping resource loading to both processors;
it is wrong to evaluate energy efficiency of Guadalupe design
by  comparing  the \texttt{lwip} performance on M3, as reported in Section \ref{sec:case}, 
with the current Linux TCP/IP performance on A9.
Unlike the heavily optimized Linux network stack, our \texttt{lwip} port is not only untuned but also has various implementation-specific limitations, e.g., maximum 64 KB TCP send buffer. 
Our reported \texttt{lwip} performance should only be read as an evidence showing that resource loading can be executed well on 
weak processors, even with such a preliminary implementation.

\subsubsection{Switch Overhead}

During page opening, as system resource demand ramps up, resource loading will be switched from M3 to A9; 
the latter processor is expected to be in low-power state before the switch happens.
The switch consists of two main steps, inter-processor interrupt and power state transition, 
which take 20-30 $\mu$s and up to 2 ms, respectively \cite{lin2012hotpower}.
As the switch happens only once in opening each web page, which typically takes $\sim$2 secs in total,
we believe the overhead is acceptable.

\subsubsection{Data Sharing Overhead}
Due to OMAP4's extreme heterogeneity for energy efficiency, no hardware cache-coherence exists between A9 and M3. 
Thus, in order to make sure that A9 has a consistent view of loaded resources, 
M3 must flush its cache before A9 can start to parse any loaded resources,
an overhead that is absent in pinning resource loading on A9. 

In order to estimate an upper bound of the overhead, 
we run a micro benchmark on M3 to periodically flush its entire 32 KB cache. 
Our measurement shows that the flush operation takes M3 $\sim$3000 cycles, or 15 $\mu$s, to complete. 
Again, as flushing happens only once in opening each web page, 
we think the overhead is acceptable.


\section{Related Work}\label{sec:related}
With the advent of heterogeneous hardware architecture, OS support for heterogeneous hardware management emerges. 
The Linux community is working toward supporting the heterogeneous multi-processor aware scheduler~\cite{linuxbiglittle} for ARM big.LITTLE~\cite{greenhalgh2011arm} architecture. 
The authors of PTask~\cite{rossbach2011sosp} propose new OS abstractions to manage GPUs as shared compute resources instead of I/O devices. 
Renderscript~\cite{renderscript} enables native Android applications to run general computation operations with automatic parallelization across all available processor cores, including GPU and DSP. 

However, even with OS support, exploiting hardware heterogeneity still requires the knowledge of web application, thus can hardly be done by OS.
Therefore, we propose that browser should manage the heterogeneous hardware directly, which is essentially an application of two generic principles: 1) the end-to-end argument~\cite{saltzer1984tocs} and 2) that the browser can be treated as the library OS~\cite{engler1995exokernel} for web applications. 

Guadalupe is the first to exploit hardware heterogeneity for web applications. 
It is designed to run on commodity OS, e.g., Android, on which native applications and web applications coexist. 
A browser OS can potentially manage hardware processing units for web applications.
For example,  ServiceOS~\cite{moshchuk2010serviceos} brings OS functions into the browser and provides secure access control and fair resource sharing mechanisms for using system resources.
But it has more freedom to modify both the browser and the kernel, since web applications are the only applications in the system. 
Moreover, previous browser OSes, e.g., Chromium OS~\cite{chromium}, IBOS~\cite{tang2010osdi} and ServiceOS~\cite{moshchuk2010serviceos}, were not designed to utilize multiple hardware processing units because the heterogeneous architecture just starts to become pervasive on mobile SoCs in recent days.

Gibraltar~\cite{lin2012webapp} abstracts the interaction between web pages and hardware components with a client-server model.
It uses AJAX as the hardware access protocol and its main focus is on I/O devices such as sensors.
Guadalupe utilizes the existing hardware abstraction to access hardware functionality, but it is able to select the most suitable hardware candidate based on the run-time web application state.

The W3C's Device APIs Working Group~\cite{w3cdap} produces standardized APIs for web applications to access device hardware,
in order to hide platform-specific hardware from web applications. 
Sharing a similar goal, Guadalupe provides the policy and mechanism to select the most suitable heterogeneous hardware processing unit, and thus hiding them from applications.

Application hints and profiling have been widely explored, e.g., for file buffer cache management~\cite{patterson1995sosp} and power management~\cite{anand2004mobisys, hotta2006ipdps, ioannou2011pact, magklis2003isca}. 
Based on hints or profiling results, the system can predict application behaviors and thus optimize for them. 
Generally, producing application hints requires extra development efforts and profiling requires training period before making good prediction. Fortunately, neither of them are necessary to Guadalupe, as Guadalupe is able to gain sufficient knowledge of applications by making sense of their current state.

Some researchers~\cite{badea2010hotpar, mai2012hotpar, meyerovich2010www} have sought to parallelize the browser.
They extract parallel tasks from the browser and execute them on homogeneous multi-core system.
Guadalupe is orthogonal to their work and can take their parallel tasks as new mapping pods for heterogeneous resources.

\section{Concluding Remarks}\label{sec:conclusion}
Guadalupe is the first endeavor to exploit the emerging hardware heterogeneity for web applications.
The design utilizes the heterogeneous processing units transparently.
It provides static mapping between the mapping pod and hardware specialization, and enables the browser to choose and switch among hardware processing units at run time based on web application state.
We demonstrate the benefit of Guadalupe design through the prototype browser implementation for resource loading and rendering.
The design not only makes more hardware resources available, but also improves energy proportionality.
More importantly, Guadalupe design opens the door to all kinds of browser services, that can potentially take advantage of the heterogeneous architecture for better performance and efficiency.

\bibliographystyle{plain}
\bibliography{./abr,./browser,./browsersecurity}

\begin{thebibliography}{10}

\bibitem{alexa}
{Alexa}.
\newblock The top 500 sites on the web.
\newblock \url{http://www.alexa.com/topsites}.

\bibitem{anand2004mobisys}
Manish Anand, Edmund~B. Nightingale, and Jason Flinn.
\newblock Ghosts in the machine: interfaces for better power management.
\newblock In {\em Proc. USENIX/ACM Int. Conf. Mobile Systems, Applications, \&
  Services (MobiSys)}, 2004.

\bibitem{armca9}
{ARM}.
\newblock Cortex-a9 processorr.
\newblock \url{http://www.arm.com/products/processors/cortex-a/cortex-a9.php}.

\bibitem{armcm3}
{ARM}.
\newblock An introduction to the arm cortex-m3 processor.
\newblock \url{http://www.arm.com/files/pdf/IntroToCortex-M3.pdf}.

\bibitem{badea2010hotpar}
Carmen Badea, Mohammad~R. Haghighat, Alexandru Nicolau, and Alexander~V.
  Veidenbaum.
\newblock Towards parallelizing the layout engine of firefox.
\newblock In {\em Proc. USENIX Conf. Hot Topics in Parallelism (HotPar)}, 2010.

\bibitem{barth2008security}
A.~Barth, C.~Jackson, C.~Reis, and TGC Team.
\newblock The security architecture of the chromium browser, 2008.

\bibitem{beverloo2012chrome}
P.~Beverloo.
\newblock {Bringing Google Chrome to Android}.
\newblock \url{http://peter.sh/2012/02/bringing-google-chrome-to-android/.},
  2012.

\bibitem{dunkels2001lwip}
A.~Dunkels.
\newblock Design and implementation of the lwip tcp/ip stack.
\newblock {\em Swedish Institute of Computer Science}, 2:77, 2001.

\bibitem{engler1995exokernel}
D.R. Engler, M.F. Kaashoek, et~al.
\newblock Exokernel: An operating system architecture for application-level
  resource management.
\newblock In {\em ACM SIGOPS Operating Systems Review}, volume~29, pages
  251--266, 1995.

\bibitem{chromeandroid}
{Google}.
\newblock Chrome for {Android} devices.
\newblock \url{www.google.com/chrome/android}.

\bibitem{chromemobile}
{Google}.
\newblock Chrome mobile {FAQ}.
\newblock \url{https://developers.google.com/chrome/mobile/docs/faq}.

\bibitem{webreplay}
{Google}.
\newblock Web page replay.
\newblock \url{http://code.google.com/p/web-page-replay/}.

\bibitem{renderscript}
{Google Renderscript}.
\newblock \url{http://developer.android.com/guide/topics/renderscript}.

\bibitem{greenhalgh2011arm}
P.~Greenhalgh.
\newblock {Big.LITTLE Processing with ARM Cortex-A15 \& Cortex-A7}.
\newblock 2011.

\bibitem{hameed2010understanding}
R.~Hameed, W.~Qadeer, M.~Wachs, O.~Azizi, A.~Solomatnikov, B.C. Lee,
  S.~Richardson, C.~Kozyrakis, and M.~Horowitz.
\newblock Understanding sources of inefficiency in general-purpose chips.
\newblock {\em ACM SIGARCH-Computer Architecture News}, 38(3):37, 2010.

\bibitem{hotta2006ipdps}
Yoshihiko Hotta, Mitsuhisa Sato, Hideaki Kimura, Satoshi Matsuoka, Taisuke
  Boku, and Daisuke Takahashi.
\newblock Profile-based optimization of power performance by using dynamic
  voltage scaling on a pc cluster.
\newblock In {\em Proc. Int. Conf. Parallel and distributed processing
  (IPDPS)}, 2006.

\bibitem{huang2012mobisys}
J.~Huang, F.~Qian, A.~Gerber, Z.M. Mao, S.~Sen, and O.~Spatscheck.
\newblock A close examination of performance and power characteristics of 4g
  lte networks.
\newblock In {\em Proc. USENIX/ACM Int. Conf. Mobile Systems, Applications, \&
  Services (MobiSys)}, pages 225--238. ACM, 2012.

\bibitem{pvrtune}
{Imagination Technologies}.
\newblock {PVRTune}.
\newblock \url{http://www.imgtec.com/powervr/insider/powervr-pvrtune.asp}.

\bibitem{ioannou2011pact}
N.~Ioannou, M.~Kauschke, M.~Gries, and M.~Cintra.
\newblock Phase-based application-driven hierarchical power management on the
  single-chip cloud computer.
\newblock In {\em Proc. Int. Conf. Parallel Architectures and Compilation
  Techniques (PACT)}, 2011.

\bibitem{OpenGL}
{Khronos Group}.
\newblock {OpenGL}.
\newblock \url{http://www.opengl.org}.

\bibitem{lin2012asplos}
F.X. Lin, Z.~Wang, R.~LiKamWa, and L.~Zhong.
\newblock Reflex: using low-power processors in smartphones without knowing
  them.
\newblock In {\em Proc. ACM Int. Conf. Architectural Support for Programming
  Languages \& Operating Systems}, 2012.

\bibitem{lin2012hotpower}
F.X. Lin, Z.~Wang, and L.~Zhong.
\newblock Supporting distributed execution of smartphone workloads on loosely
  coupled heterogeneous processors.
\newblock In {\em Proc. Workshp. Power-Aware Computing and Systems (HotPower)},
  2012.

\bibitem{lin2012webapp}
K.~Lin, D.C.J. Mickens, L.Z.F. Zhao, and J.~Qiu.
\newblock Gibraltar: exposing hardware devices to web pages using {AJAX}.
\newblock In {\em Proc. USENIX Conf. Web Application Development}, 2012.

\bibitem{linuxbiglittle}
{LWN.net}.
\newblock Linux support for {ARM big.LITTLE}.
\newblock \url{http://lwn.net/Articles/481055/}.

\bibitem{magklis2003isca}
Grigorios Magklis, Michael~L. Scott, Greg Semeraro, David~H. Albonesi, and
  Steven Dropsho.
\newblock Profile-based dynamic voltage and frequency scaling for a multiple
  clock domain microprocessor.
\newblock In {\em Proc. Int. Symp. Computer Architecture (ISCA)}, 2003.

\bibitem{mai2012hotpar}
H.~Mai, S.~Tang, S.T. King, C.~Cascaval, and P.~Montesinos.
\newblock A case for parallelizing web pages.
\newblock In {\em Proc. USENIX Conf. Hot Topics in Parallelism (HotPar)}, 2012.

\bibitem{meyerovich2010www}
Leo~A. Meyerovich and Rastislav Bodik.
\newblock Fast and parallel webpage layout.
\newblock In {\em Proc. Int. Conf. World Wide Web (WWW)}, 2010.

\bibitem{microsoftsurface}
{Microsoft}.
\newblock {Surface} tablet.
\newblock \url{http://www.microsoft.com/Surface/en-US/surface-with-windows-rt.}

\bibitem{moshchuk2010serviceos}
A.~Moshchuk and H.J. Wang.
\newblock Resource management for web applications in serviceos.
\newblock Technical report, Microsoft Research, 2010.

\bibitem{patterson1995sosp}
R.~H. Patterson, G.~A. Gibson, E.~Ginting, D.~Stodolsky, and J.~Zelenka.
\newblock Informed prefetching and caching.
\newblock In {\em Proc. ACM Symp. Operating Systems Principles}, 1995.

\bibitem{rossbach2011sosp}
C.J. Rossbach, J.~Currey, M.~Silberstein, B.~Ray, and E.~Witchel.
\newblock {PTask}: operating system abstractions to manage {GPUs} as compute
  devices.
\newblock In {\em Proc. ACM Symp. Operating Systems Principles}, pages
  233--248, 2011.

\bibitem{saltzer1984tocs}
J.H. Saltzer, D.P. Reed, and D.D. Clark.
\newblock End-to-end arguments in system design.
\newblock {\em ACM Transactions on Computer Systems (TOCS)}, 2(4):277--288,
  1984.

\bibitem{samsungnote}
{Samsung}.
\newblock {Galaxy Note 10.1}.
\newblock
  \url{http://www.samsung.com/global/microsite/galaxynote/note_10.1/benefits.h%
tml}.

\bibitem{tang2010osdi}
S.~Tang, H.~Mai, and S.T. King.
\newblock Trust and protection in the illinois browser operating system.
\newblock In {\em Proc. USENIX Conf. Operating systems design and
  implementation (OSDI)}, pages 1--8, 2010.

\bibitem{tiblaze}
{Texas Instruments}.
\newblock {Blaze Tablet}.
\newblock \url{http://omappedia.org/wiki/OMAP4_BlazeTablet}.

\bibitem{omap4470}
{Texas Instruments}.
\newblock {OMAP4470}.
\newblock \url{http://www.ti.com/product/OMAP4470}.

\bibitem{omapconf}
{Texas Instruments}.
\newblock {OMAPCONF}.
\newblock \url{https://github.com/omapconf/omapconf}.

\bibitem{omap4}
{Texas Instruments}.
\newblock {OMAP4} applications processor: Technical reference manual.
\newblock \url{http://www.ti.com/product/OMAP4470}, 2010.

\bibitem{tibltsville}
{Texas Instruments BLTsville}.
\newblock \url{http://graphics.github.com/bltsville/}.

\bibitem{chromium}
{The Chromium Projects}.
\newblock Chromium {OS}.
\newblock \url{http://www.chromium.org/chromium-os}.

\bibitem{chromiumgpu}
{The Chromium Projects}.
\newblock {GPU} accelerated compositing in {Chrome}.
\newblock
  \url{http://dev.chromium.org/developers/design-documents/gpu-accelerated-com%
positing-in-chrome}.

\bibitem{TIPET}
{TI}.
\newblock Power estimation tool.
\newblock \url{http://www.ti.com/tool/powerest/}.

\bibitem{w3cdap}
{W3C Device {APIs} Working Group}.
\newblock \url{http://www.w3.org/2009/dap}.

\bibitem{wang2012www}
Zhen Wang, Felix~Xiaozhu Lin, Lin Zhong, and Mansoor Chishtie.
\newblock How far can client-only solutions go for mobile browser speed?
\newblock In {\em Proc. Int. Conf. World Wide Web (WWW)}, 2012.

\bibitem{framerate}
{Wikipedia}.
\newblock {Frame rate}.
\newblock \url{http://en.wikipedia.org/wiki/Frame_rate}.

\end{thebibliography}
\end{document}